\documentclass[preprint,12pt]{elsarticle}

\usepackage{amssymb}

\usepackage{makecell}
\usepackage{amsmath,amssymb,amsfonts}
\usepackage{algorithmic}
\usepackage{graphicx}
\usepackage{textcomp}
\usepackage{multirow}
\usepackage{booktabs}
\usepackage{float}
\usepackage{url}
\usepackage{graphics}
\usepackage{subfigure}

\usepackage{color}

\begin{document}

\begin{frontmatter}

\bibliographystyle{elsarticle-num}

%\title{Payload Anomaly Detection Based on Deep Feature Learning \px{Deeply Detecting Anomalies in Packet Payload}}
\title{ Deep Anomaly Detection in Packet Payload}

\author[address1]{Jiaxin Liu\fnref{myfootnote}}
\author[address1,address2]{Xucheng Song\fnref{myfootnote}}
\fntext[myfootnote]{The first two authors contributed equally to this work.}

\author[address1]{Yingjie Zhou\corref{mycorrespondingauthor}}
\cortext[mycorrespondingauthor]{Corresponding author}
\ead{yjzhou@scu.edu.cn; yjzhou09@gmail.com}

\author[address1]{Xi Peng}
\author[address3]{Yanru Zhang}
\author[address1,address4]{Pei Liu}
\author[address4]{Dapeng Wu}

\address[address1]{College of Computer Science, Sichuan University, Chengdu, 610065, China}
\address[address2]{School of Information and Communication Engineering, University of Electronic Science and Technology of China, Chengdu, 611731, China}
\address[address3]{College of Computer Science and Engineering, University of Electronic Science and Technology of China, Chengdu, 611731, China}
\address[address4]{Department of Electrical and Computer Engineering, University of Florida, Gainesville, FL, 32611, USA}
\begin{abstract}
With the widespread adoption of cloud services, especially the extensive deployment of plenty of Web applications, it is important and challenging to detect anomalies from the packet payload. For example, the anomalies in the packet payload can be expressed as a number of specific strings which may cause attacks. Although some approaches have achieved remarkable progress, they are with limited applications since they are dependent on in-depth expert knowledge, e.g., signatures describing anomalies or communication protocol at the application level. Moreover, they might fail to detect the payload anomalies that have long-term dependency relationships. To overcome these limitations and adaptively detect anomalies from the packet payload, we propose a deep learning based framework which consists of two steps. First, a novel feature engineering method is proposed to obtain the block-based features via block sequence extraction and block embedding. The block-based features could encapsulate both the high-dimension information and the underlying sequential information which facilitate the anomaly detection. Second, a neural network is designed to learn the representation of packet payload based on Long Short-Term Memory (LSTM) and Convolutional Neural Networks (CNN). Furthermore, we cast the anomaly detection as a classification problem and stack a Multi-Layer Perception (MLP) on the above representation learning network to detect anomalies. Extensive experimental results on three public datasets indicate that our model could achieve a higher detection rate, while keeping a lower false positive rate compared with five state-of-the-art methods.

\end{abstract}


% \begin{highlights}
% \end{highlights}


\end{frontmatter}


\section{Introduction}
%Web application plays an important role in our information society. It is not only a convenient carrier for delivering news and exchanging information, but also offers a simple interface for users to access various services via smartphones, mobile devices, or the front ends of cloud systems. 
The rapid increase of cloud services brings remarkable convenience to our daily life and promotes the Internet economy. However, it is faced with abundant threats from malicious attackers. According to the data from the annual report of Micro Focus~\cite{MICROFOCUS2018report}, there are almost 51\% growth of disclosed vulnerabilities that are related with Web applications in 2017, and nearly 95\% of Web applications are vulnerable to sensitive data exposure, which would cause great harm to the usage of cloud services. Therefore, it is highly expected to accurately detect anomalies in network traffic. To this end, a variety of methods have been developed, which could be roughly classified into the following categories, namely, rule-based methods, flow-based methods, and packet-based methods.  

As one typical method of rule-based anomaly detection, Carmen et al.~\cite{torrano2011applying} applied feature selection called Generic-Feature-Selection to construct domain specific rules for Web application firewall. By adopting and integrating these technology~\cite{torrano2015combining,lin2004constructing}, a number of powerful tools have been developed for constructing domain specific rules from known threats, such as Suricata~\cite{suricata} and Snort~\cite{roesch1999snort}. These tools use a highly efficient engine to discover malicious traffic by comparing the extracted signatures with the predefined rules. If malicious traffic is detected, actions can be taken to protect the cloud services. Although the rules-based methods are effective for the known threats, they heavily depend on in-depth expert knowledge, e.g., signatures describing anomalies. 

Recently, some machine learning methods have been proposed to detect traffic anomaly. There are two popular directions, which use flow-based information and packet-based information to detect anomalies respectively. Flow-based anomaly detection usually treats the representation of network traffic as a type of time series \cite{zhou2019dual,zhang2018adaptive,ren2017piecewise}. \cite{xu2005profiling} used the five-tuple to construct comprehensive behavior profiles of network traffic in terms of communication patterns of end-hosts and services. The anomalies are detected by exploring the correlation between the traffic behaviors and the corresponding characteristics. \cite{da2016atlantic} presented a framework called ATLANTIC which uses similarities of flows to detect threats in traffic flows. These methods could achieve a competitive performance when the flow-based behaviors are presented. However, they do not perform well for some kinds of attacks, e.g., shell-code and SQL injection, which do not express abnormal characteristics in flow-based information.

\begin{figure}[H]
\centering
\includegraphics[width=\textwidth]{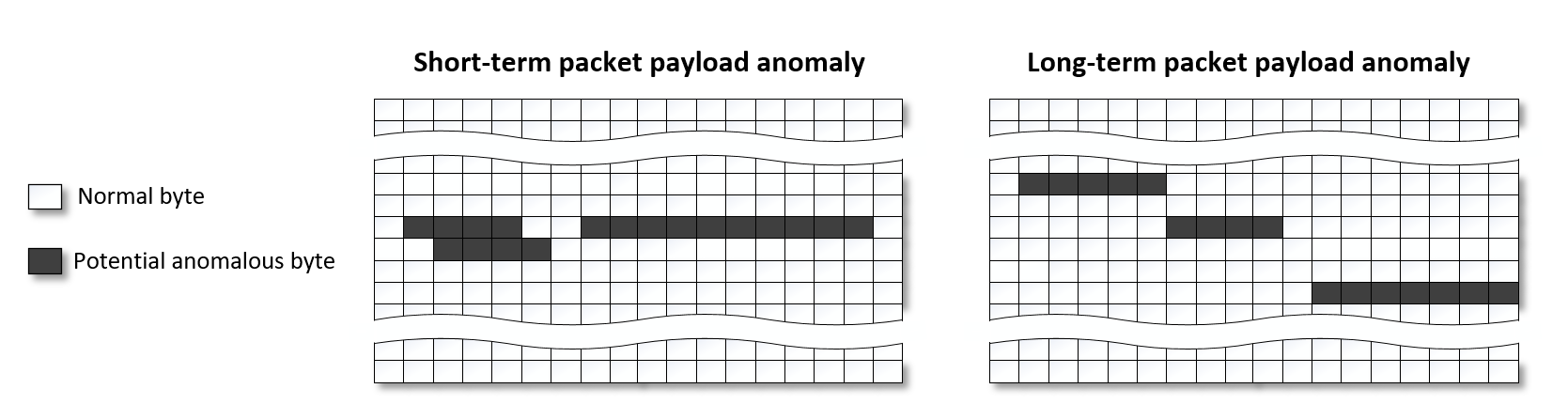} 
\centering
\caption{Two examples of the packet payload anomalies that have different distributions of anomalous bytes. For the short-term packet payload anomaly, the potential anomalous bytes are concentrated and their abnormal characteristics may be obvious, which could be detected by existing methods. In contrast, for the long-term packet payload anomaly, the potential anomalous bytes are scattered and their abnormal characteristics may not be addressed by existing works. Long-term anomalies in packet payload are more difficult to be detected than short-term ones for the existing methods.}
\label{longtermlabel}
\end{figure}

Packet-based anomaly detection can unveil anomalies by inspecting the packet payload, which refers to the user data of network packet. The objective of packet-based anomaly detection is to discover the possible attacks that have potential abnormal characteristics in the packet payload. The anomalies might appear as a number of specific strings. For example, as one of the most common attacks, i.e., SQL injection, which injects anomalous codes, such as `` ' or 1=1 - -'', into conditional statements of SQL queries to make them always be true. To detect this kind of anomalies from packet payload, a variety of methods have been proposed. The PAYL was proposed in \cite{wang2004anomalous}, which used 1-gram frequency distribution of the packet payload as features to detect network anomalies. McPAD was then proposed in \cite{perdisci2009mcpad}, which developed a modified feature extraction method for accurate anomaly detection. More recently, deep learning technology is explored for payload anomaly detection. Several literature \cite{marin2018rawpower,zhang2017a,qin2018attentional} investigated by using raw measurements to detect payload anomalies. These methods use deep learning technologies to automatically extract features from packet payload. However, the performance of payload anomaly detection is still undesirable due to the incomplete representation of features. As shown in Figure \ref{longtermlabel}, there are two types of packet payload anomalies that have different distributions of anomalous bytes. Unlike the short-term packet payload anomalies whose anomalous bytes are concentrated, the anomalous bytes for long-term are scattered and their abnormal characteristics can not be addressed by existing works. Most of the existing detection methods ignore the long-term dependency relationships among the anomalous bytes.

To tackle this, we propose a payload anomaly detection framework, which consists of two parts. The former part of the proposed framework is a feature engineering method, which consists of two steps. First, it introduces a sliding block to construct block sequences from packet payload. Second, the low-frequency items of block sequences are filtrated by a dictionary and the high-frequency items are encoded into the low-dimension embedded vectors by a self-learning block embedding layer. The proposed feature engineering method constructs the block-based features, which contain both the high-dimension information and the underlying sequential information to reveal the characteristics of payload. The latter part of the proposed framework is a detection model, which has an LSTM and CNN based neural network for learning both the potential long-term and short-term dependency relationships among the block-based features and an MLP based classifier to discover potential attacks. The major contributions of this paper could be summarized as follows:
\begin{itemize}
	\item We propose a feature engineering method that constructs block-based features of the packet payload, which could reveal the long-term dependency relationships among the anomalous bytes in packet payload. Our feature engineering method are not dependent on in-depth expert knowledge. To the best of our knowledge, this could be the first work that explores the long-term dependency relationships among the anomalous bytes for the payload anomaly detection.
	\item We design a detection model that contains an LSTM and CNN based neural network to learn both the long-term and short-term dependency relationships in the block-based features and an MLP based classifier to discover potential attacks in the packet payload.
	\item We evaluate the proposed framework that integrates the feature engineering method and the detection model by using three public datasets.
\end{itemize}

The rest of the paper is organized as follows. Section 2 introduces the related work, including the traditional technology and the deep learning technology for network anomaly detection. Section 3 presents the proposed framework, which integrates a feature engineering method and a detection model. In Section 4, we evaluate the proposed framework by using three public datasets. We conclude the paper in Section 5.

\section{Related work}

\subsection{Network Anomaly Detection}

{Network anomaly detection is a fundamental task for the quality of service (QoS) of Internet. A lot of previous work focused on the anomaly detection of low-level network flows or high-level backbone networks. To detect anomalies through flow-based information, ATLANTIC\cite{da2016atlantic} used deviations in the entropy of traffic flow tables to detect threats in traffic flows. In \cite{xu2005profiling}, K. Xu et al. detected anomalies by exploring the correlation between traffic behaviors and the corresponding characteristics in backbone networks. These methods could achieve a high detection accuracy for flow-based anomalies, but it is unlikely to detect attacks that insert anomalies in packet payloads, e.g., shell-code and SQL injection. Packet-based anomaly detection methods focus on inspecting the abnormal information in the packet payload. K. Wang et al. \cite{wang2004anomalous} proposed PAYL which uses the 1-gram frequency distribution of the payload as features to detect anomalies. R. Perdisci et al. \cite{perdisci2009mcpad} proposed McPAD to construct modified 2-gram features that contain abundant information for accurate anomaly detection. However, the accuracy of these methods heavily depend on feature construction that is complex and requires in-depth expert knowledge.}

\subsection{Deep Learning Methods for Network Anomaly Detection}

{Deep learning technology, which could automatically learns representation of data, was recently explored to address the limitations of the traditional machine learning methods. To detect the flow-based threats in network traffic, many studies investigated the power of deep learning for flow-based anomaly detection. Kim et al.\cite{kim2018web} proposed C-LSTM neural network for effectively modeling the spatial and temporal information contained in raw data to detect anomalies in traffic. Tang et al. \cite{tang2016deep} proposed a flow-based Deep Neural Network (DNN) model for intrusion detection in a software defined networking environment. To detect payload-based attacks, several detection models using the raw payload data as input have been investigated in the literature. Gonzalo et al. \cite{marin2018rawpower} applied deep CNN and LSTM neural networks for network intrusion detection with different representations of payload data. Arne et al.\cite{bochem2017streamlined} applied LSTM neural networks to learn latent characteristics of normal requests. H.Liu et al.\cite{liu2019cnn} implemented an end-to-end deep learning detection models using raw payload data. Wei et al.\cite{wang2017hast} proposed hierarchical spatial-temporal features-based intrusion detection system, which applied deep CNN to learn the low-level spatial features of network traffic and used LSTM to learn the high-level temporal features. Sheraz N.et al.\cite{naseer2018enhanced} developed several neural networks to build network anomaly detection models, including CNN, auto encoders and recurrent neural networks (RNN).}

\subsection{Summary}
The most related work to our paper is \cite{qin2018attentional}, which proposed a RNN model with the attention mechanism called ATPAD to detect anomalies in the packet payload. The ATPAD employs the word embedding and RNN to extract features, which are used at the attention calculation stage to capture the correlation between potential byte of payload and the detection results. Different from the ATPAD model, we propose a novel feature engineering method which utilizes the raw packet payload data to construct the block-based features. The block-based features contain two different kinds of information that retain both long-term and short-term dependency relationships among the packet payload. We also employs a neural network based on LSTM and CNN rather than the RNN model with the attention mechanism to capture the long-term dependency relationships among the anomalous bytes. To the best of our knowledge, our model achieves state-of-the-art performance on the CSIC 2010 dataset\cite{csic}.

\section{Proposed framework}

\begin{figure}[!h]
\centering
\includegraphics[width=\textwidth]{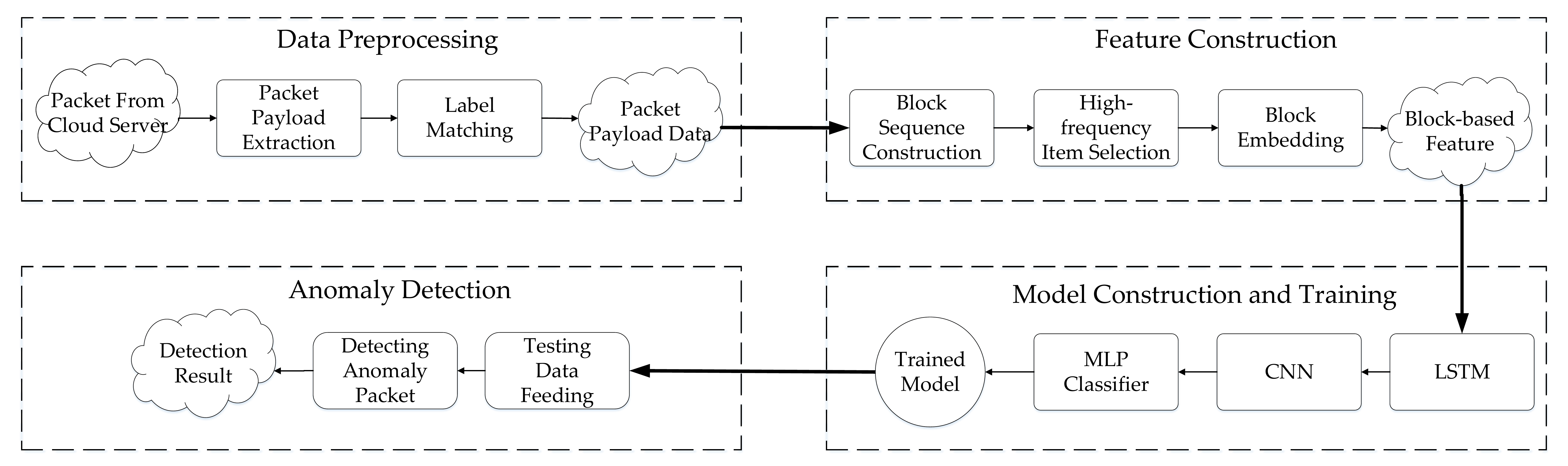} 
\centering
\caption{Overview of the proposed framework. The proposed framework contains four modules. First, the payload is extracted and labeled through a preprocessing process. Then, the block-based features are constructed for each payload. In the last two modules, a detection model based on the LSTM, CNN and MLP is designed for packet payload anomaly detection.}
\label{frameworklabel}
\end{figure}

The proposed anomaly detection framework is shown in Figure \ref{frameworklabel}. There are four modules in this framework. The first two modules make up the former part of the proposed framework, which aims to construct block-based features for efficient feature extraction. In the first module, the payload is extracted and labeled through a preprocessing process. Then, the block sequence is constructed by the sliding block and in order to remove redundant information, the high-frequency items in the block sequence are selected by a dictionary. In the block embedding process, the block-based features are constructed by encoding each item in block sequence into an embedded vector. The last two modules form the latter part of the proposed framework, which aims to adaptively detect anomalies for packet payload. Specifically, a neural network based on the LSTM and the CNN is designed to learn both the long-term and short-term dependency relationships in the block-based features and an MLP is adopted as a classifier to detect anomalies in each sample. In order to better understand how the framework works, the framework is described in details in the following subsections.

\subsection{Packet Payload Preprocessing}
The objective of the packet payload preprocessing is to extract the payload from the packet and to convert the payload into a suitable form for the following feature engineering method. The payload extraction is conducted by packet parsing based on the low-level communication protocols. The following process will try to construct efficient expression for the extracted payload. Thus, instead of employing the encoding method, e.g., popular one-hot encoding, to transform the extracted payload to an embedding vector with fixed length and possible zero padding, we directly process the whole payload to a byte stream, which is a string with variable length. The byte stream and the label with respect to the same packet make up a sample for the preprocessed packet payload data.

\subsection{Block-Based Feature Extraction}
Instead of using the payload byte stream as features, we proposed a feature engineering method to extract the block-based features which contain the high-dimension information and the underlying sequential information for anomaly detection. The block-based feature extraction has two steps, i.e., block sequence construction and block embedding. Firstly, a block sequence is constructed by using the sliding block to extract numerous items that could be considered as short subsequences. For retaining the sequential information, the items are arranged in order. Secondly, to reduce the redundant information unrelated to anomalous bytes in block sequences, the high-frequency items of each block sequence are selected by a dictionary and encoded into embedded vectors through block embedding process.
\begin{figure}
\centering
\includegraphics[width=\textwidth]{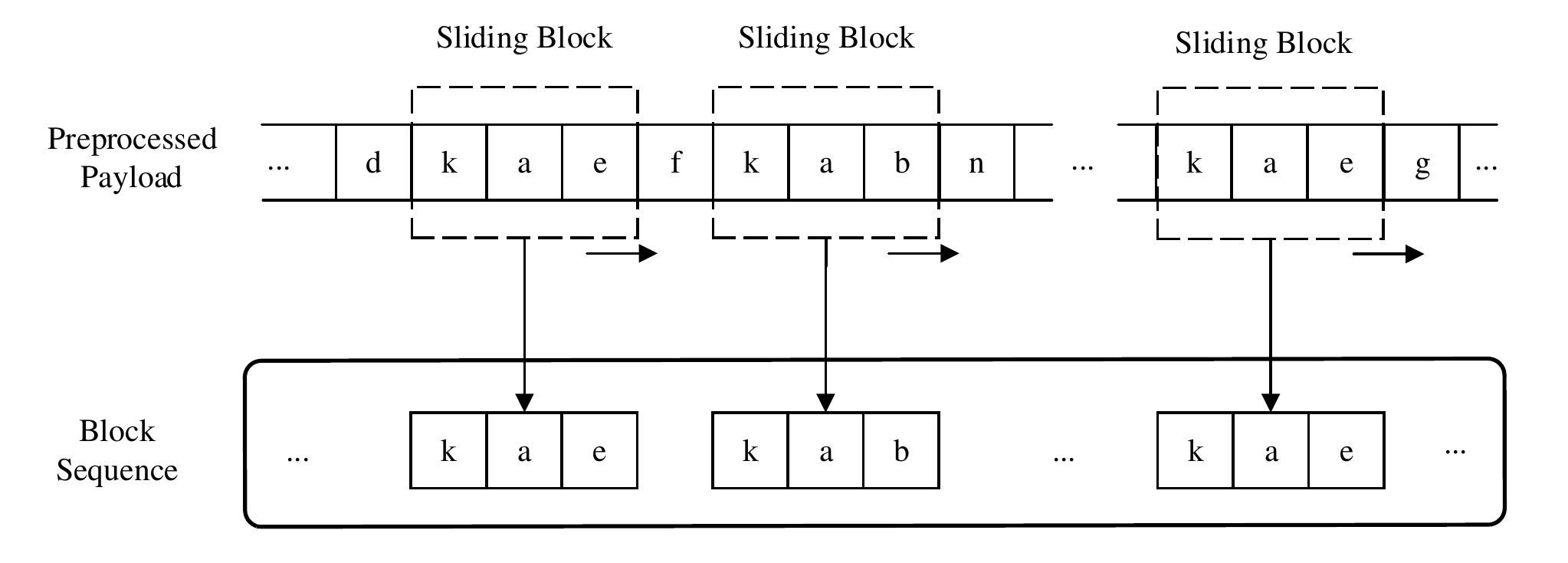}
\centering
\caption{An example for the process of block sequence construction. With a sliding block of length 3 and a fixed stride, the blocks extracted from a packet payload form a block sequence.}
\label{blocksequenceslabel}
\end{figure}

The process of block sequence extraction is shown in Figure \ref{blocksequenceslabel}. A sliding block of specific length slides on each sample consecutively. When the sliding block slides to a certain position, an item would be extracted, then the sliding block would move with a fixed stride to extract items repeatedly. Finally, the block sequence is constructed by arranging blocks in a sequence according to the order of extraction process.

\begin{figure}
\centering
\includegraphics[width=\textwidth]{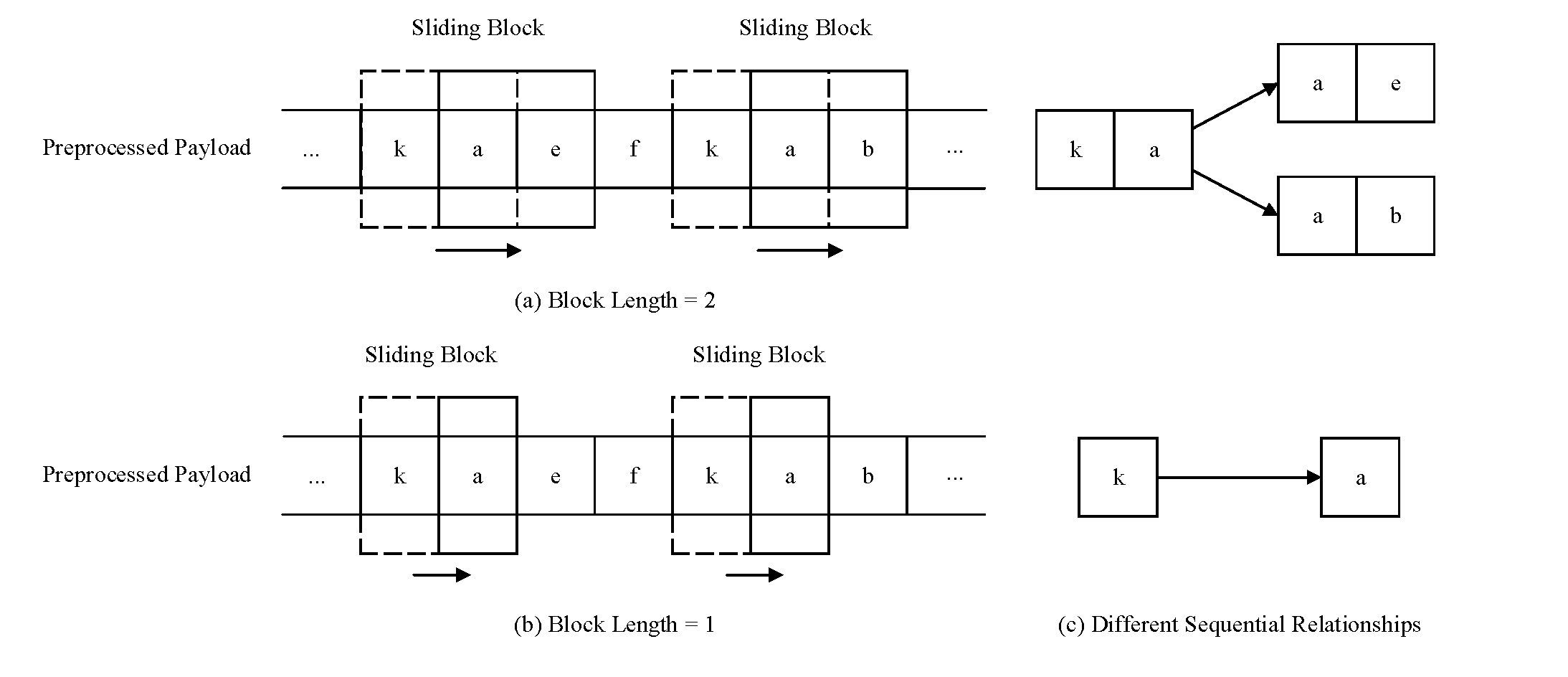}
\centering
\caption{An example of block sequence construction process and information variations using different sliding block. Using a sliding block with a fixed stride, the blocks extracted from payload could be regarded as a set of short strings. Comparing the process of sliding blocks with different lengths in the same field of the payload, the block with length 2 extracts more abundant sequential relationships than the block with length 1.}
\label{slidingwindowlabel}
\end{figure}

As mentioned above, the high-dimension information and underlying sequential information are retained in the block sequences, which are not just useful for detecting general anomalies in the payload, but also efficient for detecting anomalous bytes that have long-term dependency relationships. First of all, the high-dimension information could be considered as a kind of semantic information, which is affected by the length of sliding block. Intuitively, the longer the sliding block is, the more high-dimension information the item contains. As is shown in Figure \ref{slidingwindowlabel}(a)\&(b), for the same part of the packet payload, when the block length equals to 2, the items $ka$, $ae$, $ef$ are extracted by the sliding block. They have more information than single character $k$, $a$, $e$, $f$ that are extracted when the block length equals to 1. However, when the length of the sliding block is too long, the extracted features would contain a mixture of normal information and abnormal information, which might confuse the learning process for anomaly detection. Thus, a suitable length of the sliding block should be chosen.

Moreover, as the length of sliding block increases, the block sequence could contain more abundant sequential relationships. To be specific, under the ASCII extended $256$ standard, there are about $256^{2 n}$ possibilities of the sequential relationships between items of length $n$. As is shown in Figure \ref{slidingwindowlabel}(c), when the block length equals to 2, the item $ka$ has 2 different sequential relationships, i.e., $ka$$\longrightarrow$$ae$ and $ka$$\longrightarrow$$ab$. When the block length equals to 1, the item $k$ only contains the sequential relationship $k$$\longrightarrow$$a$. Furthermore, the expression of both the short-term and long-term dependency relationships in the block sequence is enhanced as the block length increases. This will benefit practical payload anomaly detection, especially for those that have long-term dependency relationships, such as the Union Query Attack\cite{halfond2006classification}.

The high-frequency items in the block sequence would be selected by a dictionary for the reason that there are plenty of redundant information unrelated to the anomalous bytes. In the extraction process of the sliding block, a dictionary is constructed to record the frequency of occurrence for each item and a threshold is set to limit the number of high-frequency items in the dictionary. By using the dictionary, each high-frequency item in the block sequence is selected and rearranged in the original order. Finally, each sample would be reconstructed into a sequence of selected items, which represent the significant information of each sample.

Furthermore, the high-frequency items in the block sequence are encoded by block embedding layer in order to make a better expression of the high-dimension information and underlying sequential information. One-hot encoding does not work for this task, because it could not represent the similarity between different items and with the increase of number of items, it is faced with the curse of dimensionality. Inspired from distributed representation\cite{hinton1986learning}, the items in each block sequence are encoded into low-dimension embedded vectors by a self-learning block embedding layer. The block-based features are constructed by concatenating all the vectors in order.

The proposed feature engineering method builds the block-based features, which do not rely on in-depth expert knowledge, as several low-dimension embedded vectors to form a valid expression of packet payload.

\begin{figure}
\centering
\includegraphics[width=\textwidth]{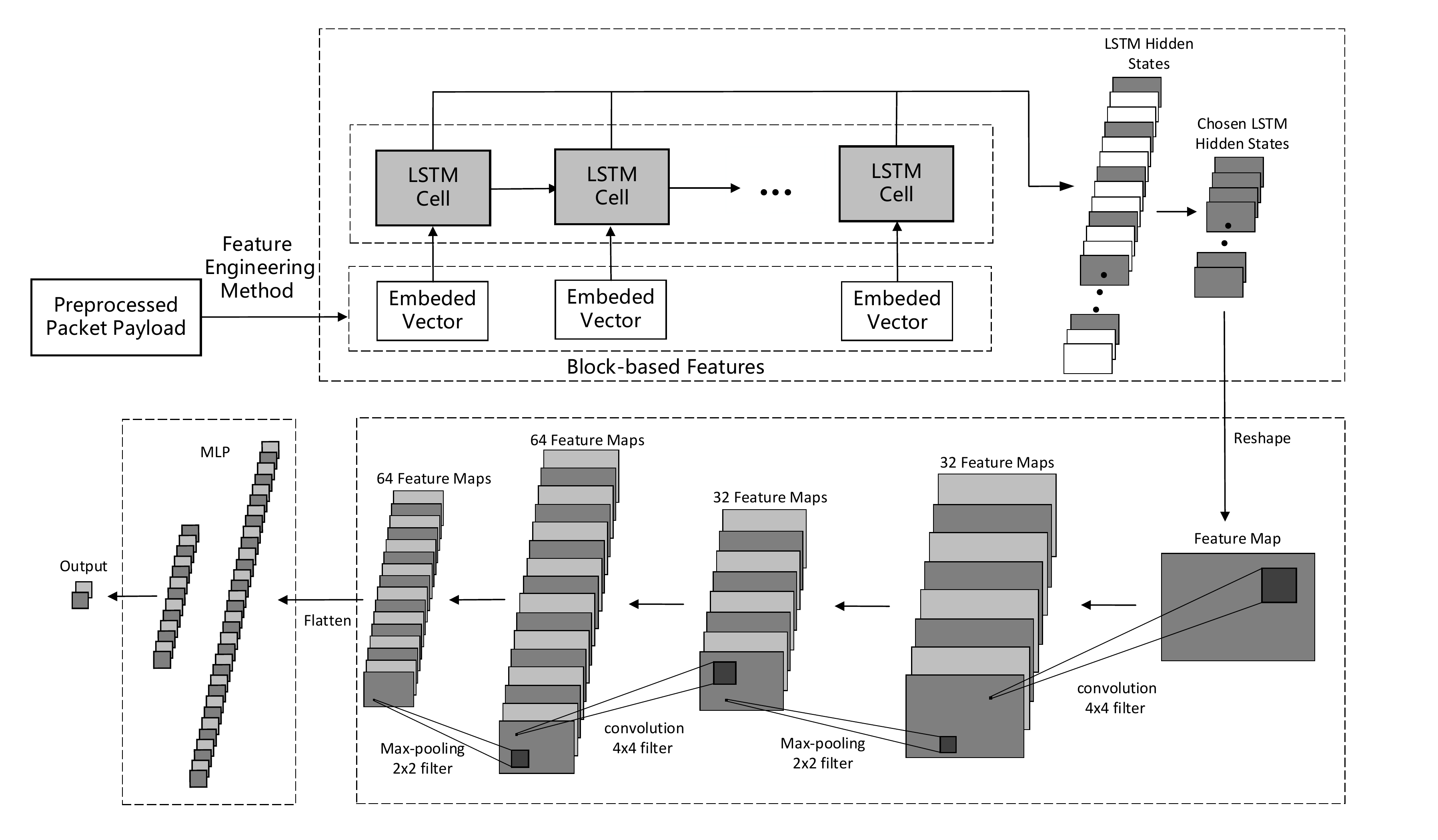}
\centering
\caption{An illustration of our proposed anomaly detection model.}
\label{totalframeworklabel}
\end{figure}

\subsection{Model Construction and Anomaly Detection}
The structure of the proposed anomaly detection model is presented in Figure \ref{totalframeworklabel}. A neural network based on LSTM and CNN is designed to learn the high-dimension information and the underlying sequential information contained in the block-based features. LSTM is used to learn the sequential dependency relationships among the block-based features, which are indicated in its hidden states of each time step. In order to learn both the long-term and the short-term dependency relationships in block-based features, we make use of the chosen LSTM hidden states in different time steps instead of only using the last hidden state that is widely adopted in classification tasks. CNN based structure is adopted to extract the local spatial information in the chosen hidden states and an MLP connected with a softmax layer is used as a classifier to detect anomalies.

In recent years, LSTM has been applied to machine translation\cite{bahdanau2014neural}, speech recognition\cite{graves2013speech}, and so on, for its capability of processing persistent information. Benefiting from its special memory cell structure and gating mechanism, it solves the exploding and vanishing gradient problems, which enable the efficient learning for long sequences. Therefore, LSTM is adopted for purpose of learning the long-term dependency relationships in the block-based features. In the proposed detection model, we employ LSTM to learn the relationships in the block-based features, i.e., the constructed features of our feature engineering method. At each time step, an embedded vector $v$ of block-based features is fed into the LSTM. The LSTM updates its cell state $c_{t}$ and outputs the current hidden state $h_{t}$ according to the previous hidden state $h_{t-1}$ and the current input $v_{t}$ through its inner non-linear operations. For the output of the LSTM at each time step $t$, the hidden state $h_{t}$ is calculated as follows\cite{Zhou2015A}:

\begin{equation}
f_{t}=\sigma\left(W_{f} \cdot\left[h_{t-1}, v_{t}\right]+b_{f}\right)
\end{equation}
\begin{equation}
i_{t}=\sigma\left(W_{i} \cdot\left[h_{t-1}, v_{t}\right]+b_{i}\right)
\end{equation}
\begin{equation}
o_{t}=\sigma\left(W_{o} \cdot\left[h_{t-1}, v_{t}\right]+b_{o}\right)
\end{equation}
\begin{equation}
c_{t}=f_{t} * c_{t-1}+i_{t} * \tanh \left(W_{c} \cdot\left[h_{t-1}, v_{t}\right]+b_{c}\right)
\end{equation}
\begin{equation}
h_{t}=o_{t} * \tanh \left(c_{t}\right)
\end{equation}

Here, the $f_{t}$, $i_{t}$ and $o_{t}$ are the forget, input and output gates respectively. They control the process for updating the LSTM hidden state. $\sigma$ is the logistic sigmoid function\cite{yin2003flexible} and the tanh is the hyperbolic tangent function\cite{xiao2005simple}. $W$ is the weight matrix and $b$ is the bias. The notations $\cdot$ and $*$ represent the Matmul product and the Hadamard product\cite{manevitz2000document} respectively.

Assume that the length of block-based features is $n$, which varies with different samples, there will be $n$ hidden states. The last hidden state of LSTM is widely adopted in classification tasks, however, it could not adequately express the long-term relationships in the block-based features. To tackle this, we choose $m$ candidates from the $n$ hidden states and these candidates are equally spaced in the ascending hidden states. This process not only preserves the long-term relationships, but also reduces the complexity of feature expression.

%\begin{figure}
%\centering
%\includegraphics[width=\textwidth]{chosen_hidden_state.eps}
%\centering
%\caption{An illustration of the LSTM structure used in the proposed detection model. By feeding the block-based features of a payload ($\mathrm{n}$ feature in total) into LSTM iteratively, the LSTM hidden state for each step is considered as a candidate to form the output of LSTM. However, only a subset of candidates, i.e., $\mathrm{m}$ candidates that are equally spaced in the ascending hidden states, is chosen to form the output of LSTM. The chosen hidden states contain information for each phase of the input, which preserve the long-term dependencies in the packet payload.}
%\label{fig:demand}
%\end{figure}

{CNN is powerful for its capability to learn spatial features and reduce feature space. Benefiting from sparse connectivity, shared weights and pooling, CNN extracts the spatial correlation information via convolution without any complex processing\cite{krizhevsky2012imagenet}. The CNN based structure in our model is used to extract high-level spatial information in the chosen hidden states. The chosen hidden states are concatenated in order and reshaped into a two-dimensional matrix. In the convolution layer, multiple convolution filters slide over the matrix to do the convolution operations, which extract the local spatial features. The learning process for CNN based structure is progressive, where the first convolution layer extracts low-level features and the next convolution layer extracts high-level features. After each convolution layer, a max-pooling layer is adopted to obtain the largest value of a small region, which preserves the important parameters and enhances the generalization ability of the model. In addition, the rectified liner unit (ReLU)\cite{nair2010rectified} is used as the activation function to add nonlinear constraint in the process. After the convolution and pooling, the spatial features of each sample are extracted and flattened into a vector which is further transmitted to the classifier.}

By casting the payload-based anomaly detection as a classification problem, an MLP is stacked on the above neural network to detect anomalies. The MLP has two layers, which would convert the flattened vector into a two-dimension vector $z$, and the softmax function maps it into a two-dimension distribution $s=\left(s_{0}, s_{1}\right)$ by Eq. (\ref{softmax}), whose values are scaled between 0 and 1, and the sum of these two values is 1. The sample would be labeled by the Eq. (\ref{addlabel}). The label 0 means the classifier judges the sample is normal, while the label 1 means the classifier judges the sample is anomalous.

\begin{equation}
s=\operatorname{Softmax}(z)
\label{softmax}
\end{equation}

\begin{equation}
label=\left\{\begin{array}{ll}{0,} & {\text { if } s_{0}<s_{1}} \\ {1,} & {\text { else }}\end{array}\right.
\label{addlabel}
\end{equation}

\subsection{Implementation}
In the proposed framework, there are four hyper-parameters needed to be set up, which includes the length of sliding block, the stride of sliding block, the number of high-frequency items in the dictionary and the number of chosen LSTM hidden states. In the experiments, the length of sliding block, the stride of sliding block, the number of high-frequency items in the dictionary, and the number of chosen states are set up as 3, 1, 15000, and 50, respectively. 

Regarding the parameters of the neural network we have designed, the hidden units of LSTM are set to 128 and the LSTM is fed with a embedded vector of 64 dimensions at each time step. Two convolution layers and two pooling layers are implemented in the CNN based structure. These two convolution layers have 32 and 64 filters, respectively. All the filters in the convolution layers are with size of $4\times4$. Each pooling layer uses max-pooling with a $2\times2$ filter.

An MLP with two layers is used as a classifier, which has 128 and 2 hidden units respectively. It finally converts the feature maps into a two-dimension vector for classification. During the training process, we set the learning rate to 0.0001 for a stable training. The dropout rate for the MLP is 0.1.

\section{Evaluation}
\label{sec:guidelines}
{In this section, we conduct various experiments to evaluate the performance and effectiveness of the proposed framework for the payload anomaly detection. We first describe the datasets and metrics used for the evaluation. Then, experiments are conducted to evaluate the performance of the proposed framework on different aspects.}
\subsection{Datasets}
We conduct experiments on three datasets to evaluate the performance of the proposed method. These three datasets contain various types of network traffic attacks. We randomly divided each dataset into three parts, the training set, the validation set and the testing set. These three sets account for 70\%, 10\% and 20\% of the total data in each dataset, respectively. The overview of three datasets are shown in Table \ref{dataset} and the detailed description of each dataset is introduced as follows.
\subsubsection{CSIC 2010} 
{The CSIC 2010 dataset\cite{csic} is developed at the Information Security Institute of Spanish Research National Council and contains thousands of Web requests which are generated automatically. The dataset consists of 72,000 normal requests and more than 25,000 anomalous requests, and all HTTP requests are marked as normal or abnormal. The CSIC 2010 dataset contains various types of Web attacks such as SQL injection, buffer overflow, information collection and so on.}
\subsubsection{CICIDS 2017}
{The CICIDS 2017 dataset\cite{cicids} contains both normal traffic and up-to-date attacks which resemble the true data. The dataset is developed by the Canadian Institute for Cyber Security. Various types of attacks, include DoS, DDos, heartbleed, web attack, infiltration and botnet, are collected in this dataset. In the experiments, we only use the traffic data collected in July 6, which contains three types of Web attacks that are related with the packet payload including Brute Force, XSS and SQL injection. }
\subsubsection{ISCX 2012}
{ISCX 2012 dataset\cite{ssh} contains network traffic which aims to describe network behaviors and intrusion patterns. This dataset contains actual traffic types such as HTTP, SMTP, SSH, IMAP, POP3, and FTP. It records packet payloads of traffic traces in the form of PCAP and the relevant profiles are publicly available for researchers. In our experiments, we use the traffic data collected in June 17th, which contains Brute Force SSH anomalies related with the packet payload.}

\begin{table}[ht]
\caption{The detailed description of three experimental dataset.}
\centering
%% \tablesize{} %% You can specify the fontsize here, e.g., \tablesize{\footnotesize}. If commented out \small will be used.
\begin{tabular}{ccccc}
\toprule
\textbf{Dataset}  &\textbf{Category}	&\textbf{Train}  &\textbf{Validation} &\textbf{Test}  \\
\midrule
\multirow{2}*{CSIC 2010}
&Anomaly &17,617 &2,439 &5,009 \\
&Normal  &50,328 &7,268 &14,404 \\
\cmidrule{1-5}
\multirow{2}*{CICIDS 2017}
&Anomaly &6,374 &951 &1,822 \\
&Normal  &14,188 &1,987 &4,053 \\
\cmidrule{1-5}
\multirow{2}*{ISCX 2012}
&Anomaly &1,618 &241 &482 \\
&Normal  &33,722 &4807 &9,616 \\

\bottomrule
\end{tabular}
\label{dataset}
\end{table}

\subsection{Performance Metric}
{In our experiments, we consider the abnormal packet payload to be a positive sample and the normal packet payload to be a negative sample. The methods performed in the experiments are evaluated on five metrics, i.e., Precision, Detection Rate (DR), False Positive Rate (FPR), Accuracy and F1-Score. These metrics are defined based on four related parameters, i.e., TP, TN, FP, FN, where TP represents the number of true positive samples, FN represents the number of false negative samples, FP represents the number of false positive samples and the TN represents the number of true negative samples. }

The definitions of the five metrics are listed as follows:

\begin{equation} { Precision }=\frac{TP}{TP
+F P}\label{eq1}\end{equation}
\begin{equation}DR=\frac{TP}{TP+FN}\label{eq2}\end{equation}
\begin{equation} FPR=\frac{FP}{FP+TN}\label{eq3}\end{equation}
\begin{equation}{Accuracy}=\frac{TP+TN}{TP+FP+FN+TN}\label{eq4}\end{equation}
\begin{equation} F_{1} \text{-score }=2\times\frac{\text{Precision}\times DR}{\text{Precision}+DR}\end{equation}

\subsection{Results and Discussion}

{In this section, we have implemented five experiments to evaluate the performance of the proposed framework in the following five aspects: }

\begin{itemize}
\item Experiment A: How is the performance of the proposed framework compared with the traditional machine learning methods and other state-of-the-art methods?
\item Experiment B: Whether the block-based features are a well expression of the packet payload characteristics?
\item Experiment C: Whether the proposed detection model can extract the long-term dependency relationships in payload anomalies?
\item Experiment D: What is the influence of the hyper-parameters in the proposed framework? 
\item Experiment E: How does the proposed model perform on other public datasets? 
\end{itemize}

\subsubsection{Experiment A: Performance compared with other methods }
\begin{figure}[htbp]
\centering
\subfigure[Detection Rates]{
\begin{minipage}[t]{0.85\linewidth}
\centering
\includegraphics[width=\textwidth ]{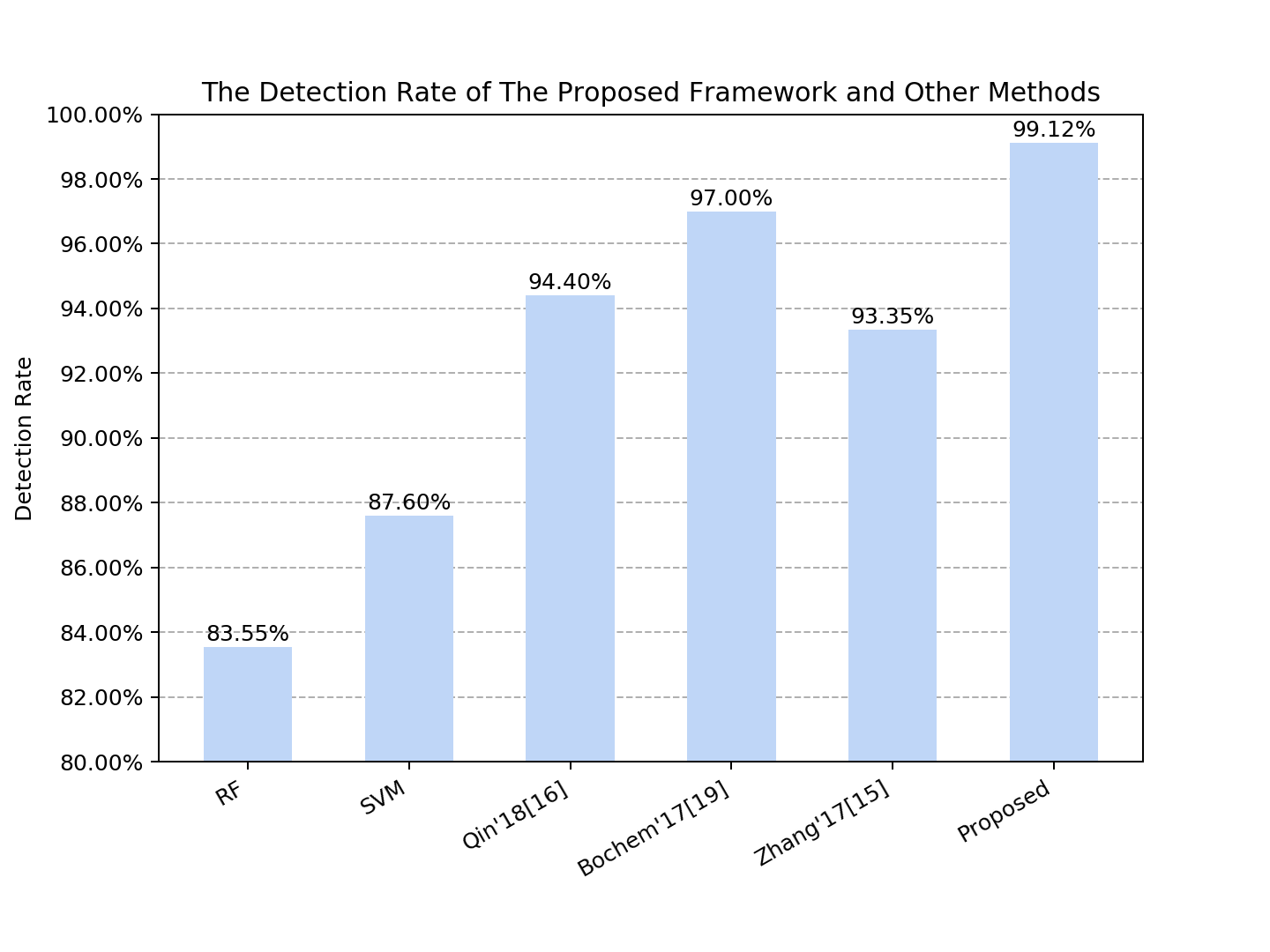}
\end{minipage}
}
\subfigure[False Positive Rates]{
\begin{minipage}[t]{0.85\linewidth}
\centering
\includegraphics[width=\textwidth]{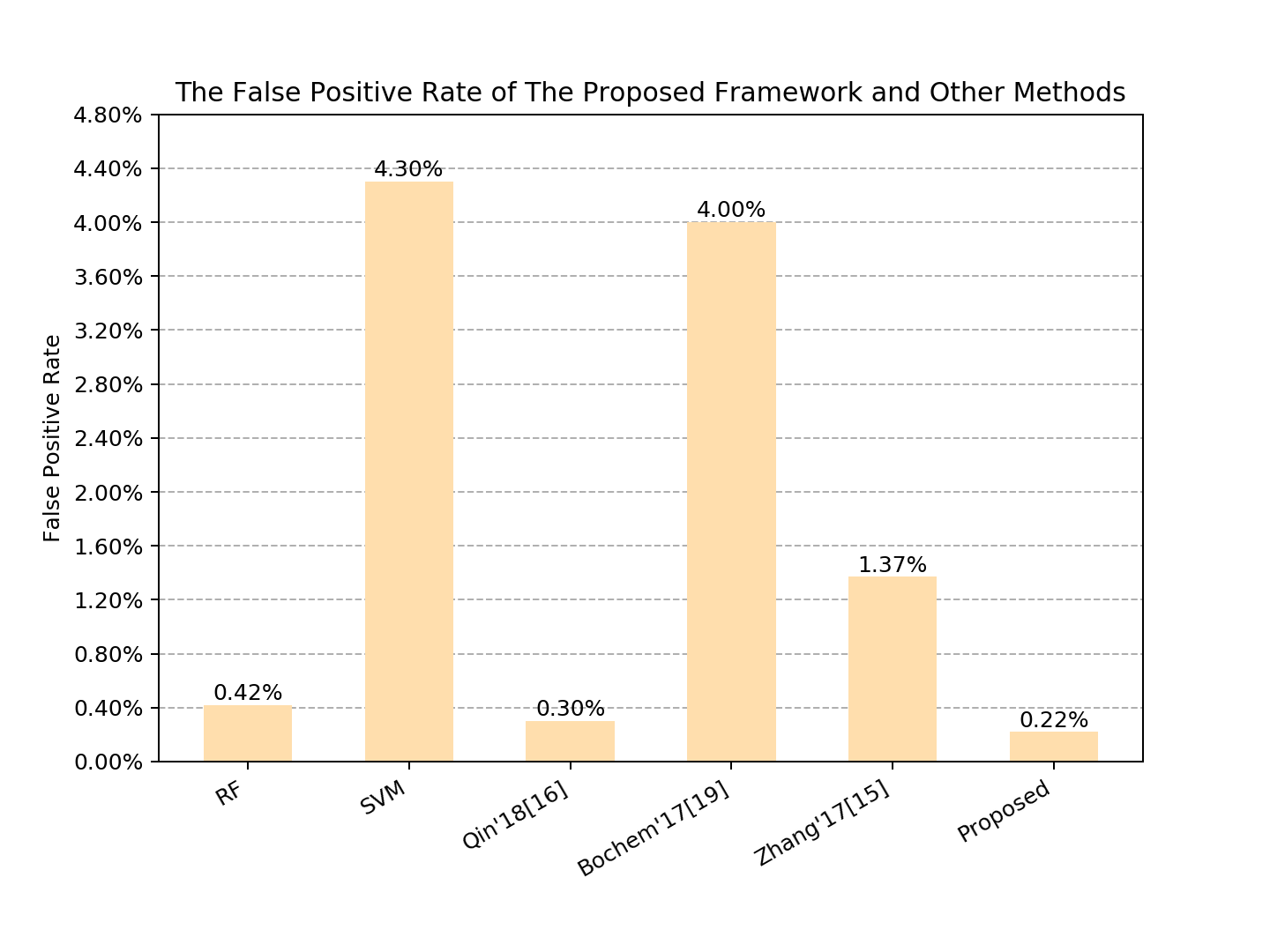}
\end{minipage}
}
\caption{The Detection Rates and False Positive Rates of the related works and our experiment.}
\label{ExperimentA}
\end{figure}

{In this experiment, we test the proposed framework on the CSIC 2010 dataset and compare the results with those of other methods. Five compared methods are involved in this experiment. Specifically, we use two classical machine learning methods and three methods recently released as the compared methods.  }
We use the scikit-learn library\cite{pedregosa2011scikit} to implement two traditional machine learning methods, support vector machine(SVM)\cite{chang2011libsvm} and random forest(RF)\cite{breiman2001random}, and test them on CSIC 2010 dataset. The other three methods include a RNN based method (Qin'18\cite{qin2018attentional}), a CNN based method (Zhang'17\cite{zhang2017a}) and a LSTM based method (Bochem'17\cite{bochem2017streamlined}). In the experiments, we simplify the http payload by ignoring the request header fields, which removes redundant information and reduces calculation complexity. The detection results of each method are listed in Figure \ref{ExperimentA}(a)\&(b), respectively.

{As shown in Figure \ref{ExperimentA}(a)\&(b), the proposed framework and three deep learning based methods outperform the two classical machine learning based methods in DR, and the proposed framework achieves the highest DR of 99.12\%. The FPR of our proposed framework is 0.22, which is lower than those of all the compared methods. To the best of our knowledge, our model achieves state-of-the-art performance on the CSIC 2010 dataset.}

\subsubsection{Experiment B: Performance analysis of the block-based features }
{In this experiment, we compare the performance of three models with or without the adoption of the block-based features. We aim to investigate whether the block-based feature engineering method can improve the anomaly detection performance in the investigated models. The first model, LSTM-CNN based model, uses the proposed network structure described in Section 3.3. The other two models are constructed by only using the LSTM network or the CNN network described in Section 3.3. We call them as the LSTM based model and the CNN based model, respectively. The “BL” prefix of a model’s name indicates the model uses the block-based features, otherwise the method uses the payload bytes stream. The CSIC 2010 dataset is used to test each model in this experiment.}
\begin{figure}[htbp]
\centering
\subfigure[Detection Rates]{
\begin{minipage}[t]{0.85\linewidth}
\centering
\includegraphics[width=\textwidth ]{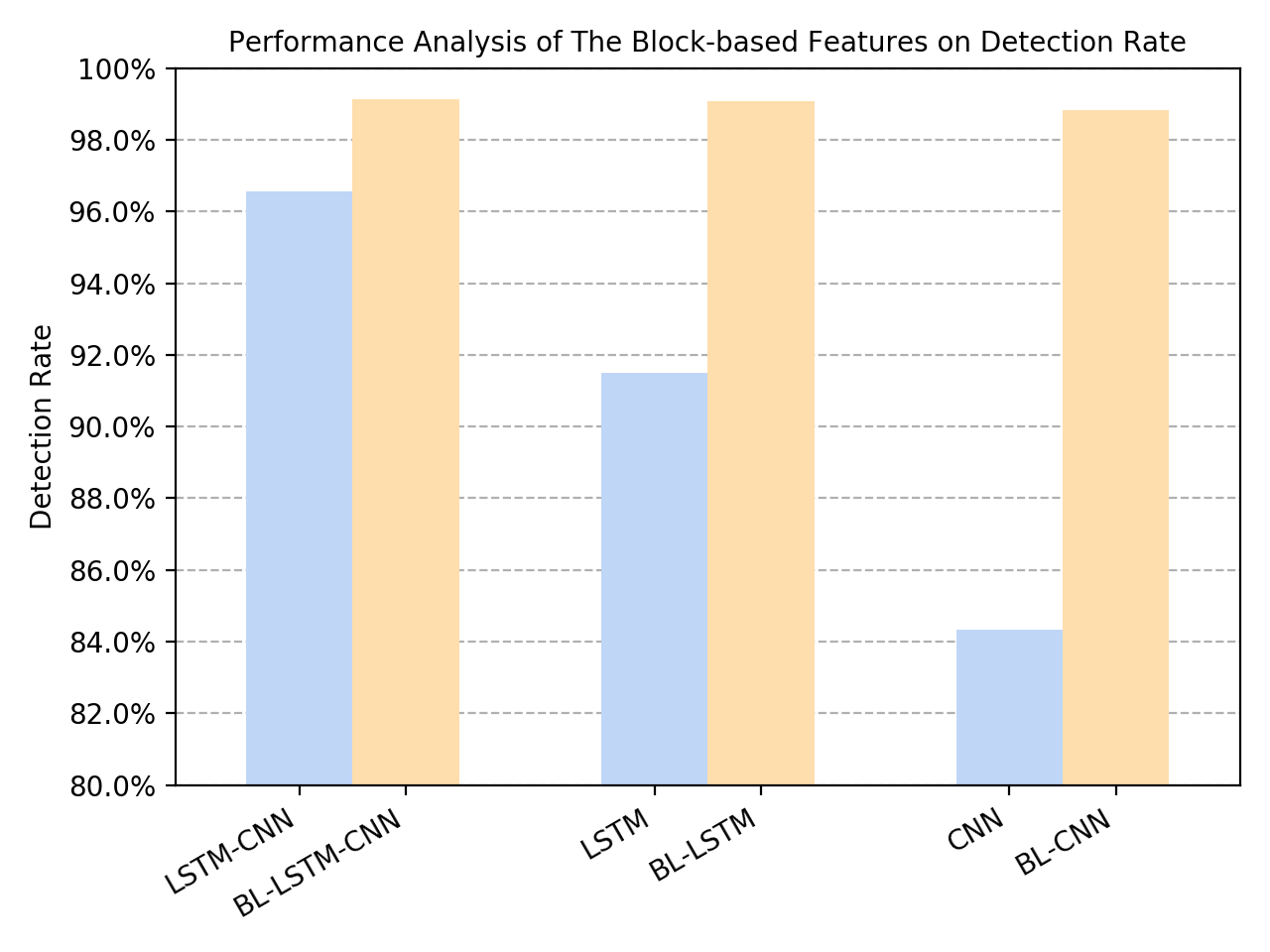}
\end{minipage}
}
\subfigure[False Positive Rates]{
\begin{minipage}[t]{0.85\linewidth}
\centering
\includegraphics[width=\textwidth]{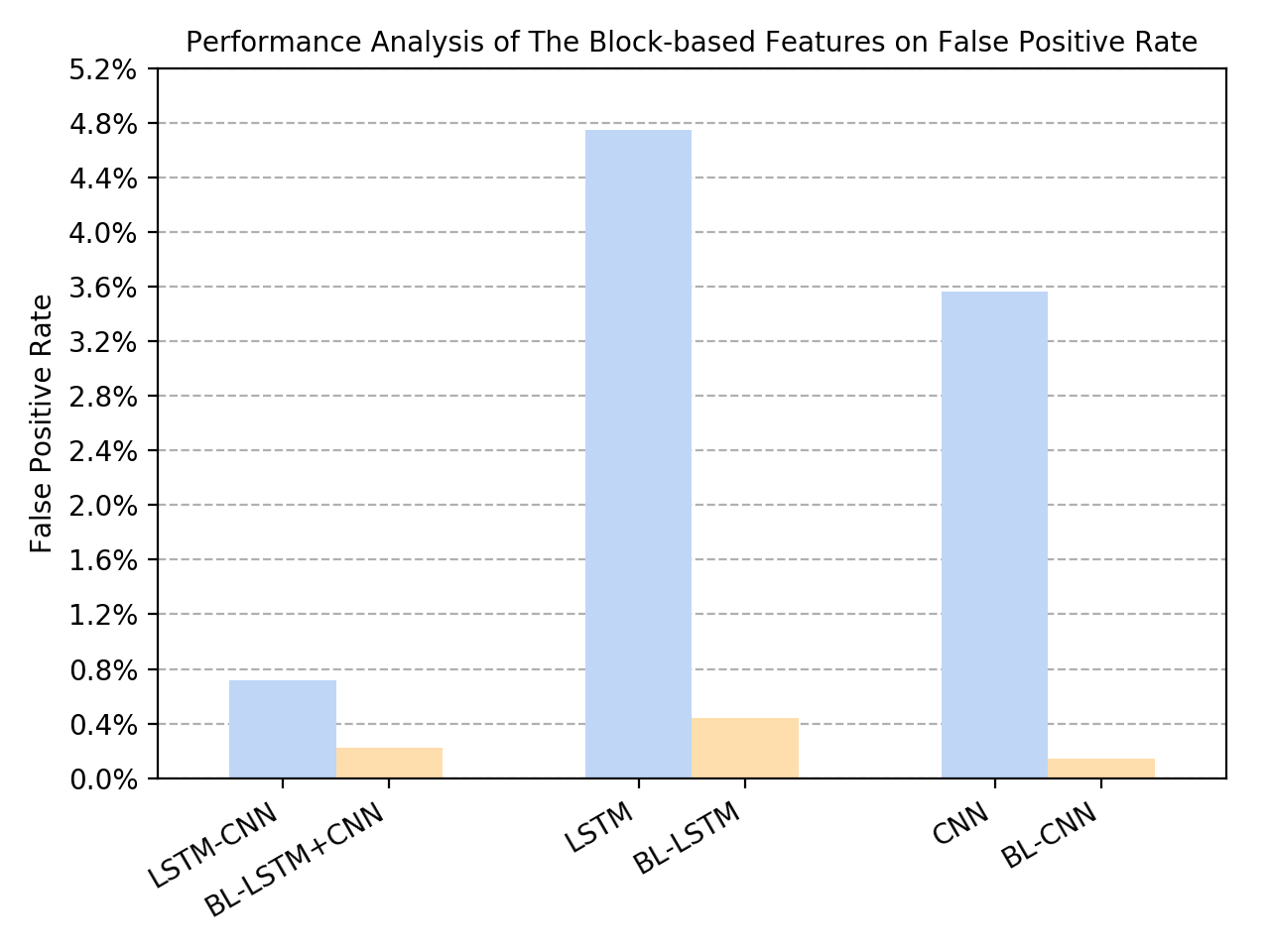}
\end{minipage}
}
\caption{Detection Rates and False Positive Rates of three models with or without the adoption of the block-based features.}
\label{ExperimentB}
\end{figure}

{Our experiment results are shown in Figure \ref{ExperimentB}(a)\&(b). When the block-based features are not used, the CNN based model achieves DR of 84.35\% and FPR of 3.56\%, the LSTM based model achieves DR of 91.5\% and FPR of 4.76\%, and the LSTM-CNN based model achieves DR of 96.57\% and FPR of 0.72\%. When the block-based features are used, the BL-CNN based model achieves DR of 98.82\% and FPR of 0.15\%, the BL-LSTM based model achieves DR of 99.08\% and FPR of 0.44\%, and the proposed model achieves DR of 99.12\% and FPR of 0.22\%. These results show that when the block-based features extraction method is applied, the DR of each model increases 14.47\%, 7.58\%, 2.55\% respectively, while the FPR of each model decreases 3.42\%, 4.31\%, 0.5\%, respectively. } 

{The above results demonstrate that the block-based features help these three models improve their detection performance on CSIC 2010 dataset. Moreover, the block-based feature extraction method could be easily combined with other anomaly detection methods and has the potential to improve their performance.}

\subsubsection{Experiment C: Performance analysis for the long-term dependency relationships in payload anomalies}
{In this experiment, we design a more challenging anomaly detection task, compared to the basic task in experiment A, to evaluate the proposed framework. The purpose of this task is to investigate whether the proposed framework has the capability of learning the long-term dependency relationships for packet payload anomalies.}

{In order to conduct the above task, we use a random-insertion method to change the sequential dependency relationships of anomalous bytes in samples of the CSIC 2010 dataset. In this method, each sample is inserted with a segment of noise, and the length of noise is 20\% of the sample length. The insertion index is random and the noise is all composed of character `0'. On one hand, after the random-insertion preprocessing, the noise is added into each sample which disrupts the original short-term dependency relationships of anomalous bytes. On the other hand, the same redundant information increases the similarity between each sample, which makes it more difficult to extract effective features and detect anomalies.}

{We calculate the performance metrics for the BL-CNN based model, the BL-LSTM model and the proposed framework. All the packet payloads are preprocessed by the random-insertion method introduced above. The experiment results are shown in Table \ref{tableExperimentC}, which indicate that the proposed framework still achieves an excellent detection performance. Compared with the results on the original CSIC 2010 dataset, the DR of the proposed framework only decreases 0.08\%, while the DR of the BL-CNN based model decreases 3.1\% and the BL-LSTM based model decreases 1.6\%. The proposed framework shows the ability to extract the long-term dependency relationships in the packet payload anomalies and still performs well on the task in this experiment. }
\begin{table}[ht]
\caption{Comparison for the results of three models in Experiment C.}
\centering
\scalebox{0.85}{
\begin{tabular}{cccccc}
\toprule
\textbf{Models} &\textbf{DR}& \textbf{FPR}& \textbf{Precision}& \textbf{$F_{1}$-score}  &\textbf{Accuracy}   \\
\midrule
BL-CNN &  95.72\% & 0.22\% & 99.34\% & 97.50\%&98.73\% \\
%\midrule
BL-LSTM &  97.47\% & 0.45\% & 98.68\% & 98.07\%&99.02\% \\
%\midrule
BL-LSTM-CNN &  \textbf{98.67\%} & \textbf{0.17\%} &\textbf{99.52}\% &\textbf{99.29\%} &\textbf{99.53\%}\\
\bottomrule
\end{tabular}}
\label{tableExperimentC}
\end{table}

\subsubsection{Experiment D: Influence of the hyper-parameters on the proposed framework}
{In this experiment, we attempt to evaluate the influence of four different hyper-parameters on the proposed framework. We evaluate one hyper-parameter each time and the other hyper-parameters are the same as those set in Experiment A. The experiment is also performed on the CSIC 2010 dataset.}
{The four parameters that we evaluate include the length of sliding block, the stride of sliding block, the number of high-frequency items in dictionary and the number of chosen LSTM hidden state. The length of sliding block is set to 1, 2, 3, 4 and 5, respectively. The block sliding length is set to 1, 2 and 3, respectively. For the evaluation of the number of high frequency items in dictionary, it is set to 5000, 10000, 15000 and 20000 each time. The number of chosen LSTM hidden state we tested includes 5, 20, 50 and 100.}

{The influence of the hyper-parameters on the proposed framework is shown in Figures \ref{ExperimentD1}\&\ref{ExperimentD2}. When the length of sliding block is too small, the information contained in the block-based features is limited. If it is too large, both normal and abnormal information might be mixed in the block-based features, which will cause poor performance. The stride of sliding block affects the amount of information extracted from the packet payload, which should be set to 1. The number of high-frequency items in dictionary affects the amount of information of block-based features, if the number is to large, the dictionary will involve too much redundant information. However, if the number is too small, the dictionary may not contain enough valuable information. The number of chosen LSTM hidden state affects the amount of sequential feature information used for the proposed framework. The experiment results indicate the model achieves its best detection performance when the number of chosen states is 50.}

\begin{figure}[b]
\centering
\subfigure[]{
\begin{minipage}{0.45\linewidth}
\centering
\includegraphics[width=\textwidth ]{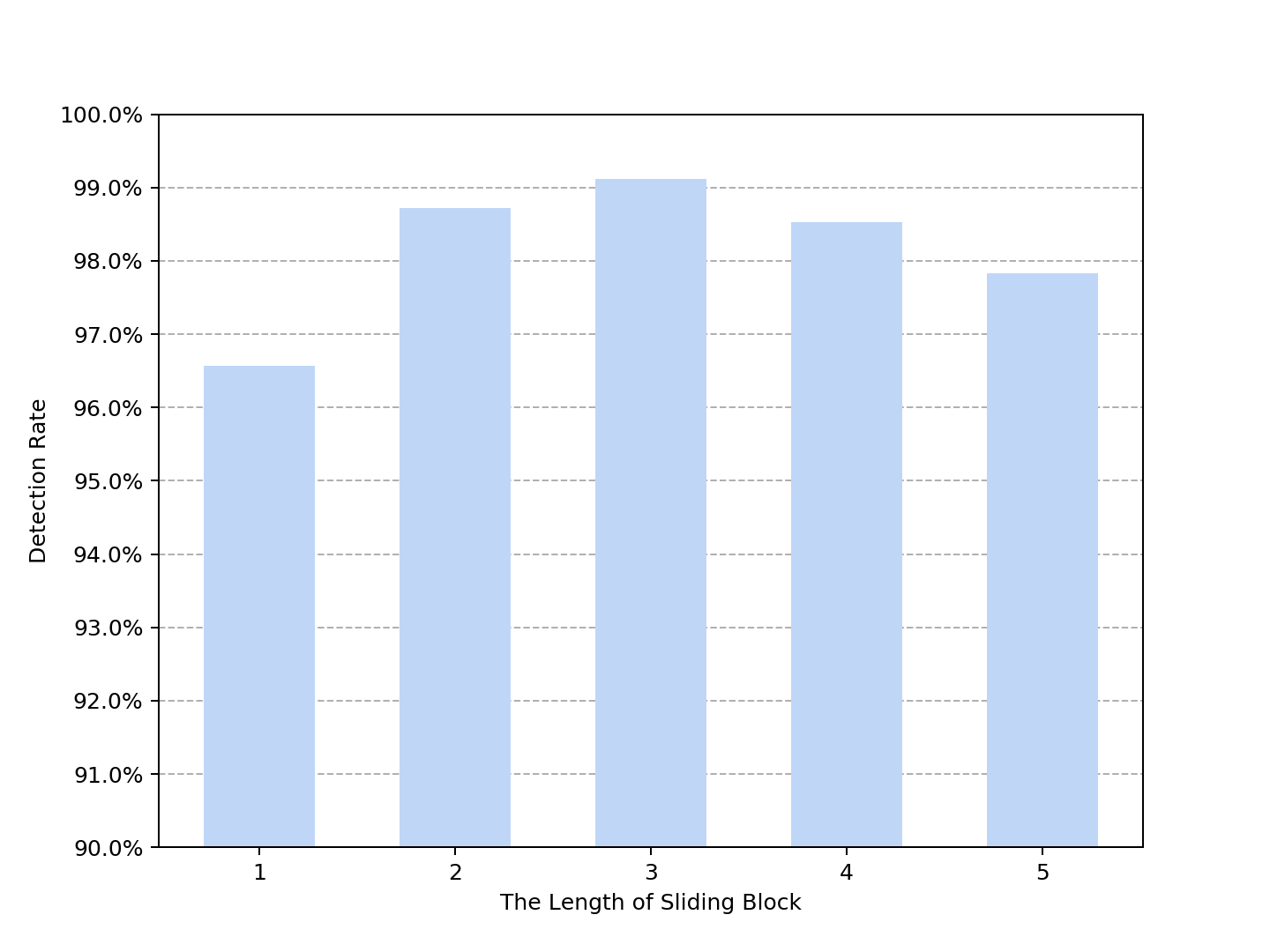}
\end{minipage}
}
\subfigure[]{
\begin{minipage}{0.45\linewidth}
\centering
\includegraphics[width=\textwidth ]{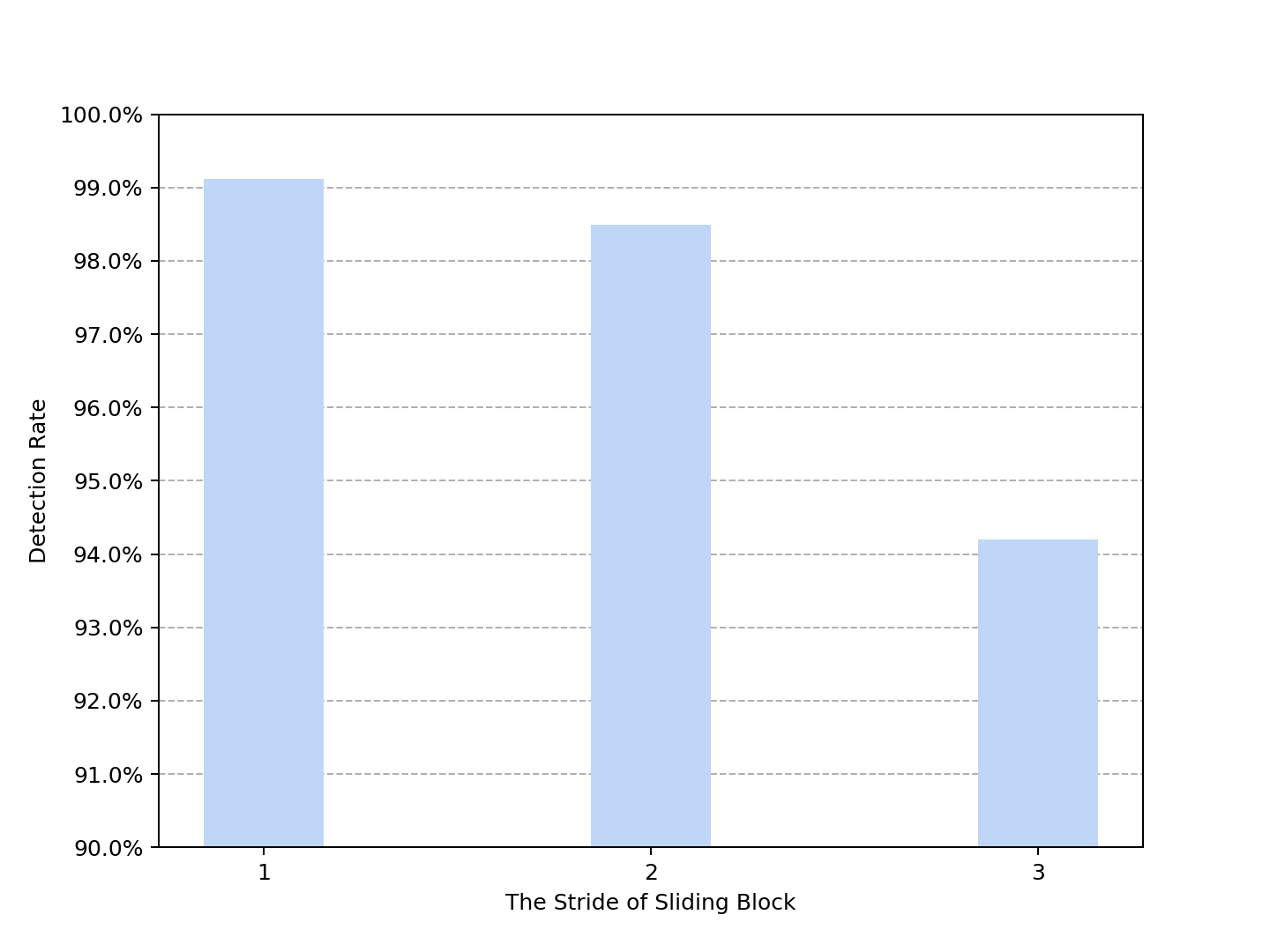}
\end{minipage}
}
\end{figure}
\addtocounter{figure}{-1}
\begin{figure}
\centering
\addtocounter{figure}{1}
\subfigure[]{
\begin{minipage}{0.45\linewidth}
\centering
\includegraphics[width=\textwidth ]{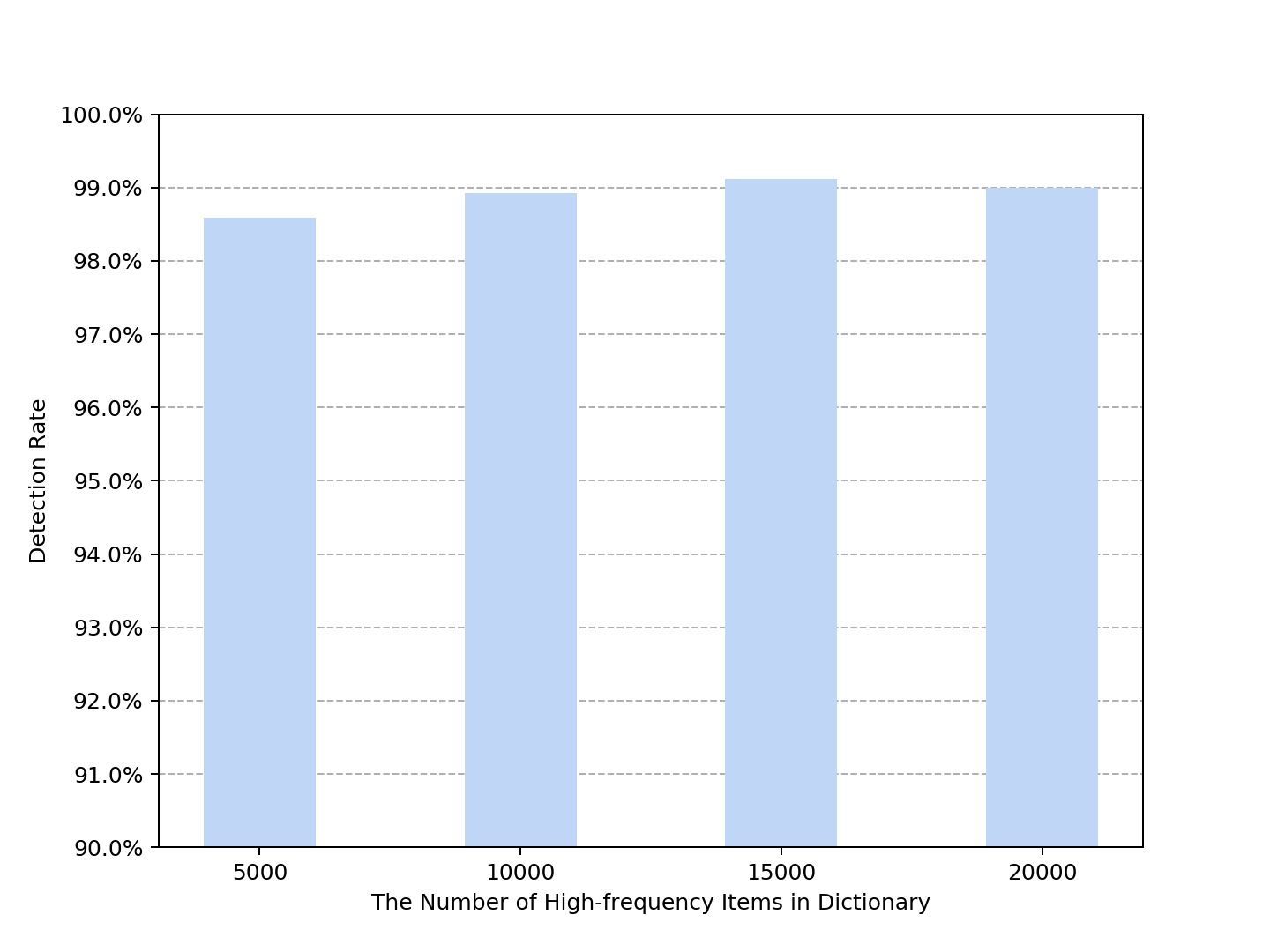}
\end{minipage}
}
\subfigure[]{
\begin{minipage}{0.45\linewidth}
\centering
\includegraphics[width=\textwidth]{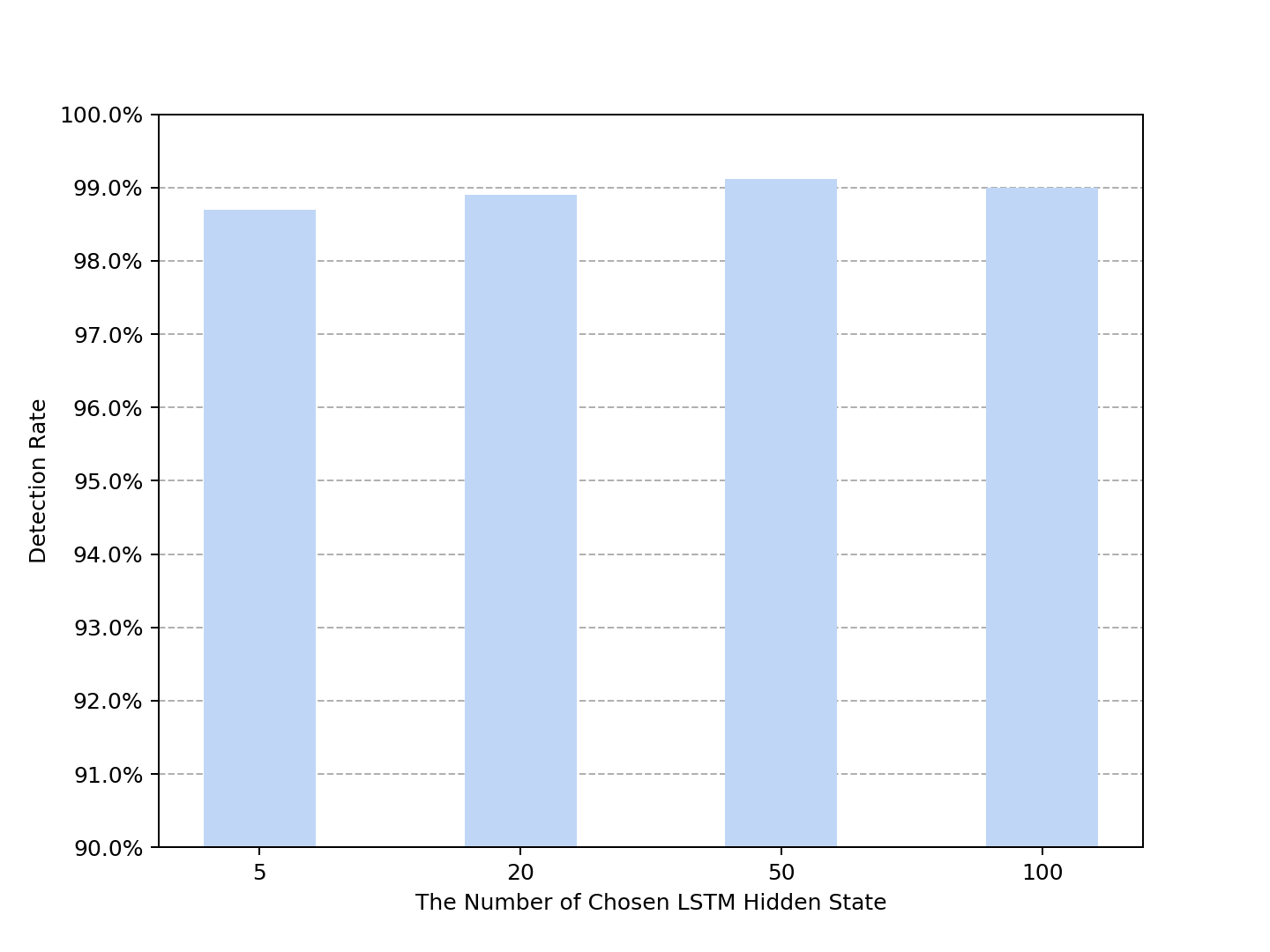}
\end{minipage}
}
\caption{The influence of model parameters on DR.}
\label{ExperimentD1}
\end{figure}
%%%%%%%%%%%%%%%%%%%fpr
\begin{figure}
\centering
\subfigure[]{
\begin{minipage}{0.45\linewidth}
\centering
\includegraphics[width=\textwidth ]{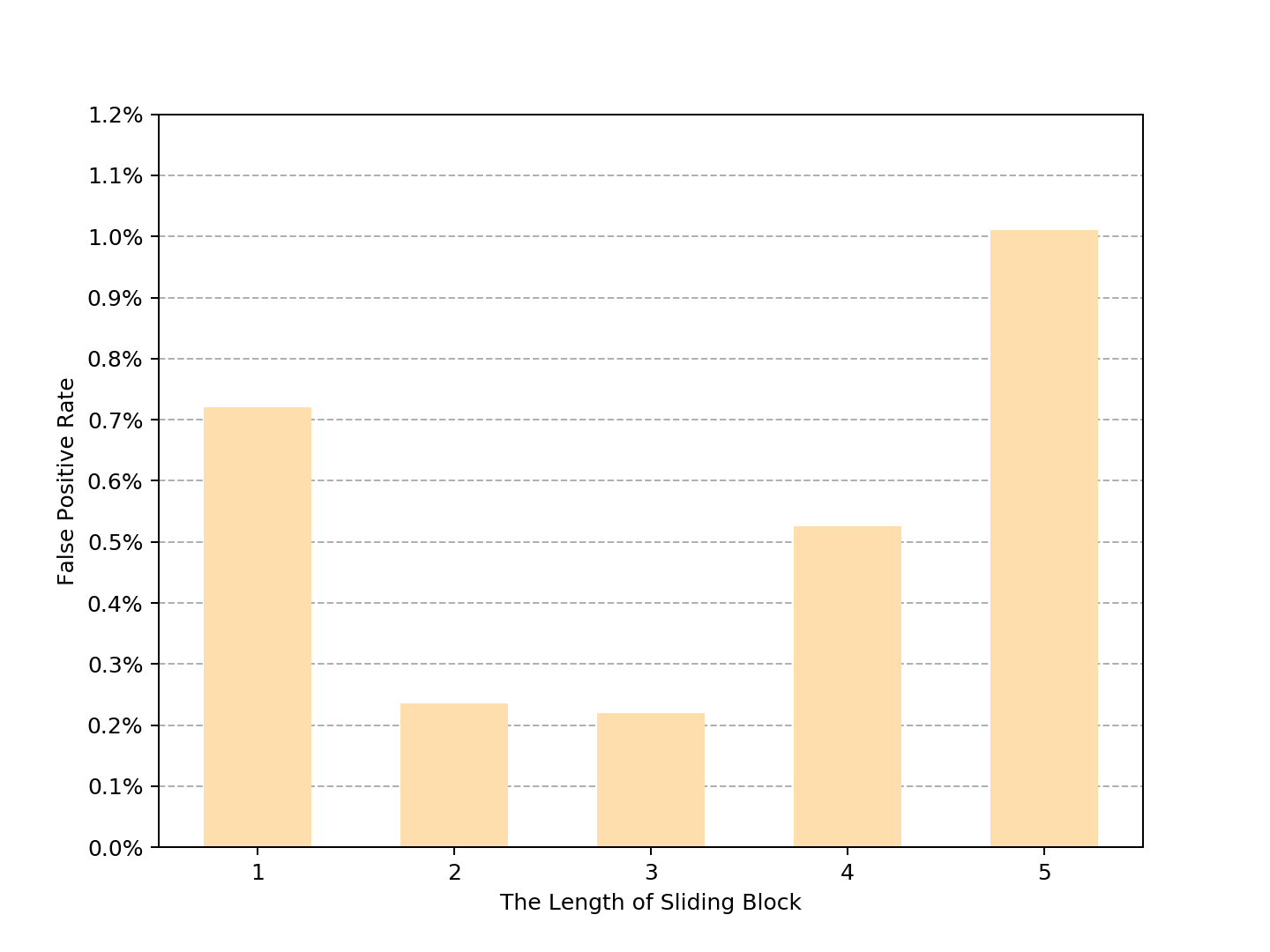}
\end{minipage}
}
\subfigure[]{
\begin{minipage}{0.45\linewidth}
\centering
\includegraphics[width=\textwidth ]{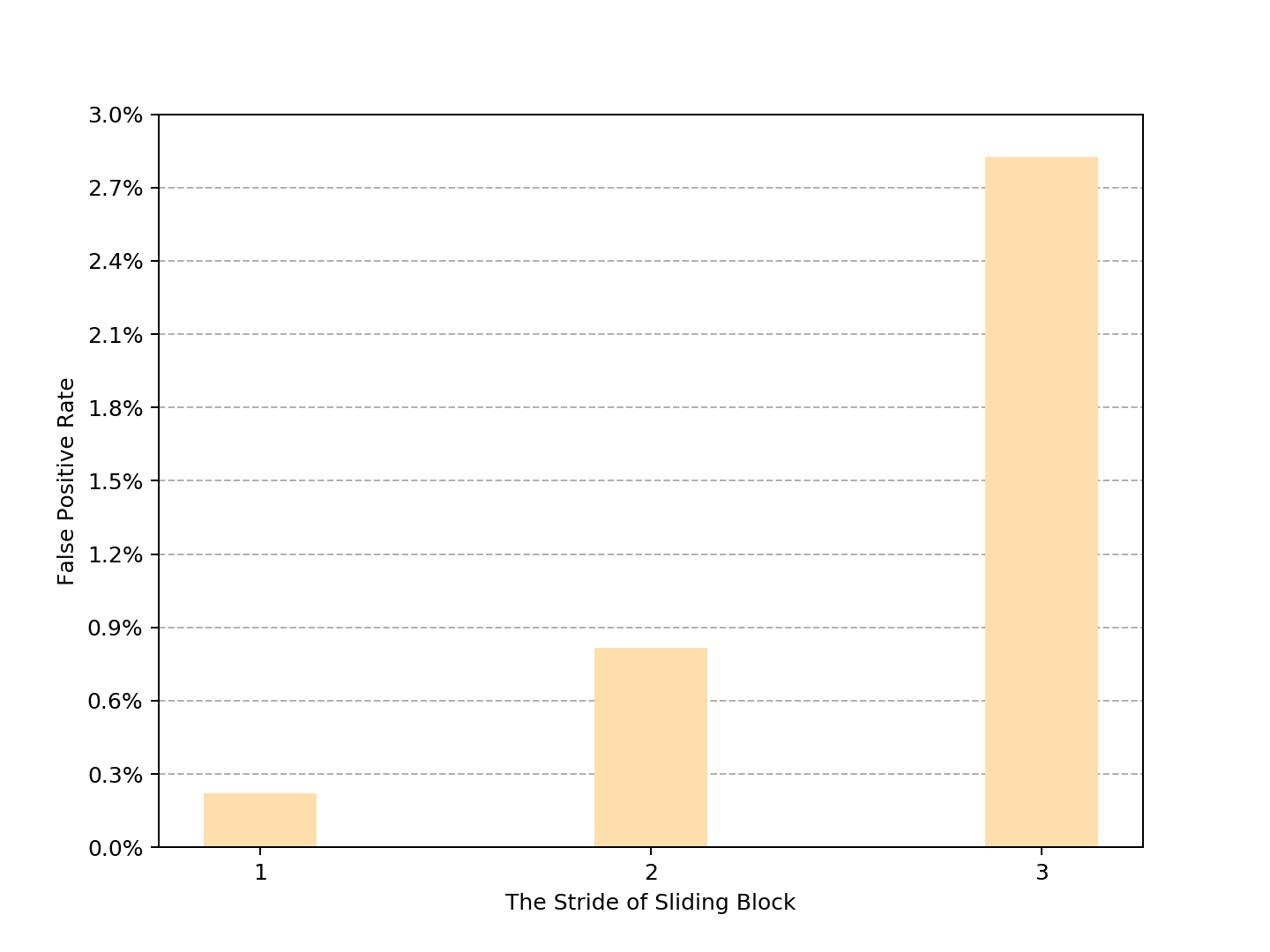}
\end{minipage}
}
\end{figure}
\addtocounter{figure}{-1}
\begin{figure}
\centering
\addtocounter{figure}{1}
\subfigure[]{
\begin{minipage}{0.45\linewidth}
\centering
\includegraphics[width=\textwidth ]{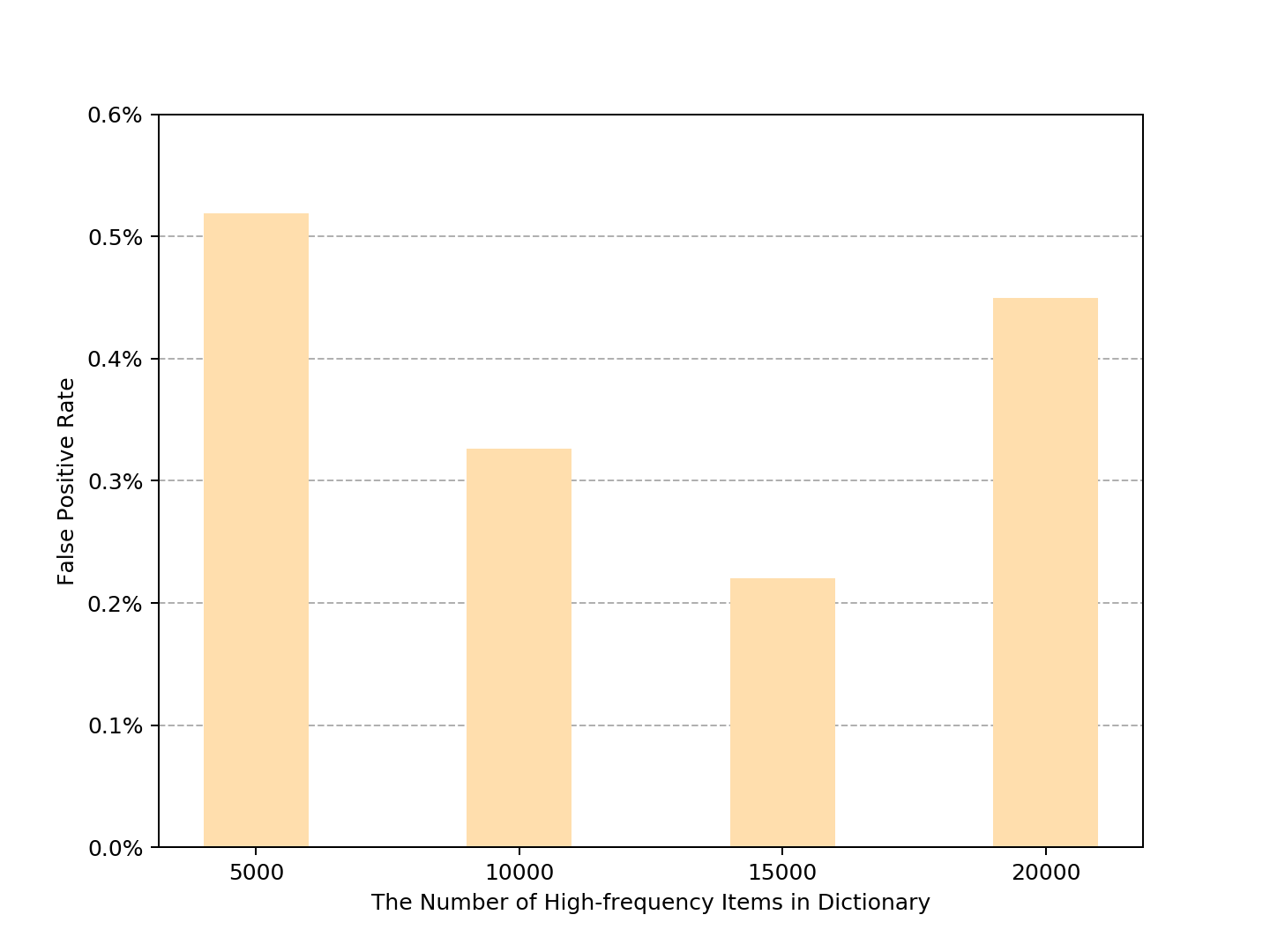}
\end{minipage}
}
\subfigure[]{
\begin{minipage}{0.45\linewidth}
\centering
\includegraphics[width=\textwidth]{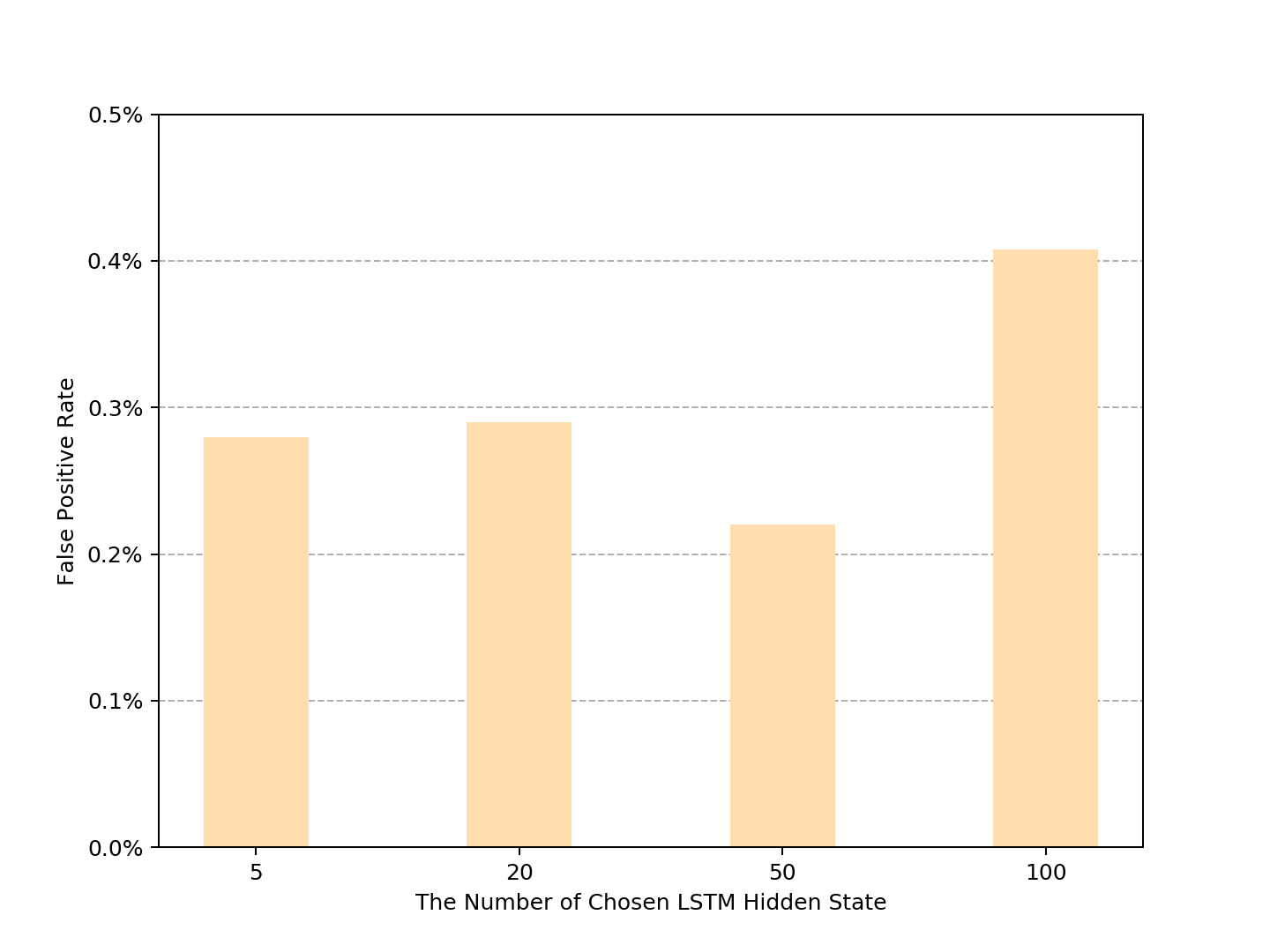}
\end{minipage}
}
\caption{The influence of model parameters on FPR.}
\label{ExperimentD2}
\end{figure}

\subsubsection{Experiment E: Performance evaluation on other public datasets}
{In this experiment, we evaluate the performance of the proposed framework on other public datasets. We use the subset of the public dataset for evaluation, which contains attacks related with the packet payload. For the CICIDS 2017 dataset, we use the Web attack data includes Brute Force, XSS, SQL injection in one day’s record to set up the attack dataset, and use the normal HTTP traffic data in that day to conduct the normal dataset. For the ISCX 2012 dataset, we only use the Brute Force SSH attack data to form the attack dataset. The normal packet payload of SSH data in that day is used to form the normal dataset. }

{The detection results on the two datasets are shown in Table \ref{tableExperimentE}. Our proposed method achieves excellent performance on both two datasets. On the CICIDS 2017 dataset, our proposed model achieves a DR of 99.78\% and an FPR of 0.0165\%. On the other dataset, our proposed model achieves a DR of 99.17\% and an FPR of 0.332\%. }

\begin{table}[htbp]
\caption{Experimental results of proposed model on other datasets.}
\centering
\scalebox{0.85}{
\begin{tabular}{cccccc}
\toprule 
\textbf{Datasets} &\textbf{DR}& \textbf{FPR}& \textbf{Precision}& \textbf{$F_{1}$-score}  &\textbf{Accuracy}   \\
\midrule
CICIDS 2017 &  99.78\% & 0.0165\% & 99.34\% & 99.56\%&99.92\% \\
%\midrule   
ISCX 2012 &  99.17\% & 0.332\% & 98.68\% & 98.92\%&99.64\% \\
\bottomrule
\end{tabular}}
\label{tableExperimentE}
\end{table}

\section{Conclusion}
This paper proposed a payload-based anomaly detection framework to construct block-based features for efficient feature extraction and to adaptively detect anomalies. The block-based features are constructed by the former part of the proposed framework, which is a feature engineering method implemented via block sequence extraction and block embedding. The latter part of the proposed framework, i.e., the anomaly detection model, is designed to learn both the long-term and short-term dependency relationships in the block-based features and to discover potential attacks in the packet payload. Experiment results with three public datasets showed that the proposed framework could achieve a high detection rate and a low false positive rate compared with existing methods in the literature. In future work, we will consider other kinds of anomalies, e.g., anomalies in video surveillance\cite{luo2019video}, and try to explore a unified framework to detect them.





\bibliography{mybibfile}

\begin{thebibliography}{10}
\expandafter\ifx\csname url\endcsname\relax
  \def\url#1{\texttt{#1}}\fi
\expandafter\ifx\csname urlprefix\endcsname\relax\def\urlprefix{URL }\fi
\expandafter\ifx\csname href\endcsname\relax
  \def\href#1#2{#2} \def\path#1{#1}\fi

\bibitem{MICROFOCUS2018report}
Micro focus 2018 application security research update,
  \url{https://www.microfocus.com/media/report/application/_security/_research/_update/_report.pdf}.

\bibitem{torrano2011applying}
C.~Torrano-Gimenez, H.~T. Nguyen, G.~Alvarez, S.~Petrovi{\'c}, K.~Franke,
  Applying feature selection to payload-based web application firewalls, in:
  2011 Third International Workshop on Security and Communication Networks
  (IWSCN), IEEE, 2011, pp. 75--81.

\bibitem{torrano2015combining}
C.~Torrano-Gimenez, H.~T. Nguyen, G.~Alvarez, K.~Franke, Combining expert
  knowledge with automatic feature extraction for reliable web attack
  detection, Security and Communication Networks 8~(16) (2015) 2750--2767.

\bibitem{lin2004constructing}
S.-C. Lin, S.-S. Tseng, Constructing detection knowledge for ddos intrusion
  tolerance, Expert Systems with Applications 27~(3) (2004) 379--390.

\bibitem{suricata}
Suricata, \url{https://suricata-ids.org/}.

\bibitem{roesch1999snort}
Snort, \url{https://www.snort.org/}.

\bibitem{zhou2019dual}
J.~T. Zhou, H.~Zhang, D.~Jin, X.~Peng, Dual adversarial transfer for sequence
  labeling, IEEE Transactions on Pattern Analysis and Machine Intelligence
  (2019) 1--1.

\bibitem{zhang2018adaptive}
L.~Zhang, J.~Lin, R.~Karim, Adaptive kernel density-based anomaly detection for
  nonlinear systems, Knowledge-Based Systems 139 (2018) 50--63.

\bibitem{ren2017piecewise}
H.~Ren, M.~Liu, Z.~Li, W.~Pedrycz, A piecewise aggregate pattern representation
  approach for anomaly detection in time series, Knowledge-Based Systems 135
  (2017) 29--39.

\bibitem{xu2005profiling}
K.~Xu, Z.~Zhang, S.~Bhattacharyya, Profiling internet backbone traffic:
  behavior models and applications, in: ACM special interest group on data
  communication, ACM, 2005, pp. 169--180.

\bibitem{da2016atlantic}
A.~S. da~Silva, J.~A. Wickboldt, L.~Z. Granville, A.~Schaeffer-Filho, Atlantic:
  A framework for anomaly traffic detection, classification, and mitigation in
  sdn, in: NOMS 2016-2016 IEEE/IFIP Network Operations and Management
  Symposium, IEEE, 2016, pp. 27--35.

\bibitem{wang2004anomalous}
K.~Wang, S.~J. Stolfo, Anomalous payload-based network intrusion detection, in:
  International Workshop on Recent Advances in Intrusion Detection, Springer,
  2004, pp. 203--222.

\bibitem{perdisci2009mcpad}
R.~Perdisci, D.~Ariu, P.~Fogla, G.~Giacinto, W.~Lee, Mcpad: A multiple
  classifier system for accurate payload-based anomaly detection, Computer
  Networks 53~(6) (2009) 864--881.

\bibitem{marin2018rawpower}
G.~Marin, P.~Casas, G.~Capdehourat, Rawpower: Deep learning based anomaly
  detection from raw network traffic measurements, in: Proceedings of the ACM
  SIGCOMM 2018 Conference on Posters and Demos, ACM, 2018, pp. 75--77.

\bibitem{zhang2017a}
M.~Zhang, B.~Xu, S.~Bai, S.~Lu, Z.~Lin, A deep learning method to detect web
  attacks using a specially designed cnn, in: International Conference on
  Neural Information Processing, Springer, 2017, pp. 828--836.

\bibitem{qin2018attentional}
Z.-Q. Qin, X.-K. Ma, Y.-J. Wang, Attentional payload anomaly detector for web
  applications, in: International Conference on Neural Information Processing,
  Springer, 2018, pp. 588--599.

\bibitem{kim2018web}
T.-Y. Kim, S.-B. Cho, Web traffic anomaly detection using c-lstm neural
  networks, Expert Systems with Applications 106 (2018) 66--76.

\bibitem{tang2016deep}
T.~A. Tang, L.~Mhamdi, D.~McLernon, S.~A.~R. Zaidi, M.~Ghogho, Deep learning
  approach for network intrusion detection in software defined networking, in:
  2016 International Conference on Wireless Networks and Mobile Communications
  (WINCOM), IEEE, 2016, pp. 258--263.

\bibitem{bochem2017streamlined}
A.~Bochem, H.~Zhang, D.~Hogrefe, Streamlined anomaly detection in web requests
  using recurrent neural networks, in: 2017 IEEE Conference on Computer
  Communications Workshops (INFOCOM WKSHPS), IEEE, 2017, pp. 1016--1017.

\bibitem{liu2019cnn}
H.~Liu, B.~Lang, M.~Liu, H.~Yan, Cnn and rnn based payload classification
  methods for attack detection, Knowledge-Based Systems 163 (2019) 332--341.

\bibitem{wang2017hast}
W.~Wang, Y.~Sheng, J.~Wang, X.~Zeng, X.~Ye, Y.~Huang, M.~Zhu, Hast-ids:
  Learning hierarchical spatial-temporal features using deep neural networks to
  improve intrusion detection, IEEE Access 6 (2017) 1792--1806.

\bibitem{naseer2018enhanced}
S.~Naseer, Y.~Saleem, S.~Khalid, M.~K. Bashir, J.~Han, M.~M. Iqbal, K.~Han,
  Enhanced network anomaly detection based on deep neural networks, IEEE Access
  6 (2018) 48231--48246.

\bibitem{csic}
Csic 2010 http dataset, \url{https://www.isi.csic.es/dataset/}.

\bibitem{halfond2006classification}
W.~G. Halfond, J.~Viegas, A.~Orso, et~al., A classification of sql-injection
  attacks and countermeasures, in: Proceedings of the IEEE International
  Symposium on Secure Software Engineering, IEEE, 2006, pp. 13--15.

\bibitem{hinton1986learning}
A.~Paccanaro, G.~E. Hinton, Learning distributed representations of concepts
  using linear relational embedding, IEEE Transactions on Knowledge and Data
  Engineering 13~(2) (2001) 232--244.

\bibitem{bahdanau2014neural}
D.~Bahdanau, K.~Cho, Y.~Bengio, Neural machine translation by jointly learning
  to align and translate, arXiv preprint arXiv:1409.0473 (2014).

\bibitem{graves2013speech}
A.~Graves, A.-r. Mohamed, G.~Hinton, Speech recognition with deep recurrent
  neural networks, in: 2013 IEEE international conference on acoustics, speech
  and signal processing, IEEE, 2013, pp. 6645--6649.

\bibitem{Zhou2015A}
C.~Zhou, C.~Sun, Z.~Liu, F.~C.~M. Lau, A c-lstm neural network for text
  classification, Computer Science 1~(4) (2015) 39--44.

\bibitem{yin2003flexible}
X.~Yin, J.~Goudriaan, E.~A. Lantinga, J.~Vos, H.~J. Spiertz, A flexible sigmoid
  function of determinate growth, Annals of Botany 91~(3) (2003) 361--371.

\bibitem{xiao2005simple}
F.~Xiao, Y.~Honma, T.~Kono, A simple algebraic interface capturing scheme using
  hyperbolic tangent function, International Journal for Numerical Methods in
  Fluids 48~(9) (2005) 1023--1040.

\bibitem{manevitz2000document}
L.~M. Manevitz, M.~Yousef, Document classification on neural networks using
  only positive examples, in: Proceedings of the 23rd annual international ACM
  SIGIR conference on Research and development in information retrieval, ACM,
  2000, pp. 304--306.

\bibitem{krizhevsky2012imagenet}
A.~Krizhevsky, I.~Sutskever, G.~E. Hinton, Imagenet classification with deep
  convolutional neural networks, in: Advances in neural information processing
  systems, 2012, pp. 1097--1105.

\bibitem{nair2010rectified}
V.~Nair, G.~E. Hinton, Rectified linear units improve restricted boltzmann
  machines, in: Proceedings of the 27th international conference on machine
  learning, 2010, pp. 807--814.

\bibitem{cicids}
Cicids 2017 dataset, \url{https://www.unb.ca/cic/datasets/ids-2017.html}.

\bibitem{ssh}
Iscxids 2012 dataset, \url{https://www.unb.ca/cic/datasets/ids.html}.

\bibitem{pedregosa2011scikit}
F.~Pedregosa, G.~Varoquaux, A.~Gramfort, V.~Michel, B.~Thirion, O.~Grisel,
  M.~Blondel, P.~Prettenhofer, R.~J. Weiss, V.~Dubourg, et~al., Scikit-learn:
  Machine learning in python, Journal of Machine Learning Research 12 (2011)
  2825--2830.

\bibitem{chang2011libsvm}
C.-C. Chang, C.-J. Lin, Libsvm: A library for support vector machines, ACM
  Transactions on Intelligent Systems and Technology 2~(3) (2011) 27.

\bibitem{breiman2001random}
L.~Breiman, Random forests, Machine learning 45~(1) (2001) 5--32.

\bibitem{luo2019video}
W.~{Luo}, W.~{Liu}, D.~{Lian}, J.~{Tang}, L.~{Duan}, X.~{Peng}, S.~{Gao}, Video
  anomaly detection with sparse coding inspired deep neural networks, IEEE
  Transactions on Pattern Analysis and Machine Intelligence (2019) 1--1.

\end{thebibliography}




% \end{thebibliography}
\end{document}